\input{epsf.sty}

\documentstyle[aps,floats]{revtex}

\begin{document}

\draft

\title{All nonspherical perturbations of the Choptuik spacetime decay}

\author{Jos\'e M. Mart\'\i n-Garc\'\i a}
\address{Laboratorio de Astrof\'\i sica Espacial y F\'\i sica Fundamental,
Apartado~50727, 28080~Madrid, Spain}

\author{Carsten Gundlach}
\address{Max-Planck-Institut f\"ur Gravitationsphysik
(Albert-Einstein-Institut), Schlaatzweg 1, 14473 Potsdam, Germany}

\date{17 September 1998}

\maketitle


\begin{abstract}

We study the nonspherical linear perturbations of the discretely
self-similar and spherically symmetric solution for a self-gravitating
scalar field discovered by Choptuik in the context of marginal
gravitational collapse. We find that all nonspherical perturbations
decay. Therefore critical phenomena at the threshold of gravitational
collapse, originally found in spherical symmetry, will extend to (at
least slightly) nonspherical initial data.

\end{abstract}

\pacs{04.25.Dm, 04.20.Dw, 04.40.Nr, 04.70.Bw, 05.70.Jk}




\section{Introduction and summary}


We have many and powerful results about the static or stationary 
end-states of gravitational collapse. However, very little is known
in comparison about the dynamical evolution towards them. Analytical studies
are limited by the complicated non-linear nature of the
equations. Numerical studies can fill this gap if they can demonstrate
generic behavior.

Starting with the pioneering work of Choptuik \cite{Choptuik},
a number of authors
have shown that, despite the complicated nature of the equations, the
threshold of gravitational collapse is strikingly simple \cite{Greview}. 
Following the initial ideas of Evans \cite{EvCol} it has been possible 
to explain this
simplicity as the consequence of the existence of a ``critical
solution'' which acts as an intermediate attractor in phase
space. This solution has a single linearly-unstable eigenmode which
drives out every nearby solution either towards black hole formation or
dispersal, leaving flat space behind.

This body of work expands our understanding of the dynamical
process of collapse, borrowing concepts and tools from the theory of
dynamical systems. The emphasis is shifted to phase space, and within
it, to solutions with special stability characteristics:
\begin{enumerate}

\item First, we look for global attractors. The Minkowski and
Kerr-Newman solutions are the only possible end-states of collapse in
the Einstein-scalar-Maxwell system.

\item Then we  look for codimension-one attractors, which separate
phase space into basins of attraction of the global attractors. These
solutions are also very important. For example, the study of the
trajectories connecting the codimension-one attractors with the global
attractors gives us a qualitative picture of marginal collapse because
many different trajectories tend to approach them and arrive at the
attractors along them. In this terminology, Choptuik discovered the
first codimension-one attractor. For reasons still unknown, many of
the codimension-one attractors are self-similar.

\item The long-term objective is the construction of a picture of the
unfolding of trajectories in phase space. It would contain all the
dynamical information about a given system. Furthermore it is the
natural place to accommodate the zoo of special solutions we currently
know of, including naked singularities.

\end{enumerate}

In a previous paper \cite{GMG} we addressed the question of
whether the Choptuik solution was a codimension-one solution in the
system Einstein-Maxwell-charged scalar field, restricted to spherical
symmetry, and obtained an affirmative answer, which has been confirmed in
independent work \cite{HodPiran}. In this paper we address the same
question for the system Einstein-real scalar field, but this time
allowing for arbitrary small deviations from spherical symmetry, and
we obtain again an affirmative answer. The study of the
Einstein-Maxwell-charged scalar field system beyond spherical symmetry
will be reported elsewhere.

This result, together with a parallel result on the collapse of a
perfect fluid \cite{critfluid}, strongly suggests that critical
phenomena in gravitational collapse are still present in the absence
of spherical symmetry. An equally strong indication that critical phenomena
are not restricted to spherical symmetry is provided by numerical
work on the critical collapse of axisymmetric vacuum spacetimes
\cite{AbrahamsEvans}, which shows universality and scaling similar
to that of the spherical scalar field.

The plan of the article is as follows: In section II we give a
complete review of the Gerlach and Sengupta \cite{GS1,GS2} formalism
of gauge-invariant perturbations around a general spherically
symmetric spacetime (which typically contains matter and is
time-dependent). In section III we re-express these still general
tensor equations of section II in an arbitrary basis to facilitate the
study of their causal structure. The equations in this section will be
of help in any study of linear perturbations around spherical symmetry
for arbitrary matter content, in an arbitrary background coordinate
system. In section IV we specialize the formalism to massless scalar
field matter. The background solution is briefly reviewed in section
V, where we specialize to a particular basis, and choose a coordinate
system. In sections VI and VII we split the odd and even linear
perturbation equations, respectively, into evolution equations and
constraints, and identify free data. Section VIII describes our
numerical results in detail. The appendix contains a description of
our numerical methods for computing the background and then the
perturbations on it.

To summarize our main result here, all non-spherical physical
perturbations of Choptuik's solution decay, and therefore the critical
phenomena at the black hole threshold in scalar field collapse ---
universality, echoing and scaling --- are expected to persist for
initial data that deviate (slightly) from spherical
symmetry. Nevertheless, the $l=2$ even-parity perturbations decay
quite slowly, and may become apparent in non-spherical collapse
situations.


\section{Review of Gerlach and Sengupta formalism of gauge-invariant
perturbations} 


In this section we give a brief introduction to the formalism of
Gerlach and Sengupta \cite{GS1,GS2} for perturbations around the most general
spherically symmetric spacetime. Spacetime is decomposed as
$M^4=M^2\times S^2$, where $S^2$ is the two-sphere and $M^2$ is a
two-dimensional manifold with boundary. Tensor indices on $M^4$ are
Greek letters, tensor indices on $M^2$ are upper case Latin letters,
and tensor indices on $S^2$ are lower case Latin letters. We write the
general spherically symmetric metric as
\begin{equation}
g_{\mu\nu}dx^\mu dx^\nu\equiv 
g_{AB}(x^D)dx^Adx^B+r^2(x^D)\gamma_{ab}(x^d)dx^adx^b ,
\label{sphericalmetric}
\end{equation}
where $g_{AB}$ is a metric and $r$ is a scalar field on $M^2$.
$\gamma_{ab}$ is the unit Gaussian-curvature metric on $S^2$.  $r=0$
identifies the center of the spherical symmetry, where each $S^2$
degenerates to a point. $r=0$ is the boundary of $M^2$.  In the same
way we decompose the spherically symmetric stress-energy tensor:
\begin{equation}
t_{\mu\nu}dx^\mu dx^\nu 
\equiv  
t_{AB}(x^D)dx^Adx^B+Q(x^D)r^2(x^D)\gamma_{ab}(x^d) dx^ad x^b .
\label{sphericalstress-energy}
\end{equation}
For simplifying the field equations, it is useful to introduce a
vector and a scalar on $M^2$ derived from the scalar $r$:
\begin{eqnarray}
v_A&\equiv &\frac{r_{|A}}{r} \label{v_A} , \\
V_0&\equiv &-r^{-2}+2{v_A}^{|A}+3v_Av^A .
\end{eqnarray}
We distinguish covariant derivatives on $M^4$, $M^2$ and $S^2$:
\begin{equation}
g_{\mu\nu;\lambda}\equiv 0 , \ \ \ \ g_{AB|C}\equiv 0 , 
\ \ \ \ \gamma_{ab:c}\equiv 0 .
\end{equation}
We shall also need the covariantly constant unit antisymmetric tensors
with respect to $g_{AB}$ and $g_{ab}$, which we call $\epsilon_{AB}$
and $\epsilon_{ab}$.

The Einstein equations in spherical symmetry are
\begin{eqnarray}
G_{AB}=-2(v_{A|B}+v_Av_B)+g_{AB}V_0 &=&8\pi  t_{AB}
\label{backgroundEinsteinAB} , \\
\frac{1}{2}{G_a}^a=-{\cal R}+{v_A}^{|A}+v_Av^A&=&8\pi  Q , 
\label{backgroundEinsteinaa}
\end{eqnarray}
where ${G_a}^a$ denotes the partial trace over $G_{\mu\nu}$.  ${\cal
R}\equiv \frac{1}{2} R^{(2)A}_A$ is the Gaussian-curvature scalar of
$M^2$. The four-dimensional Ricci scalar is $R=2({\cal R}-V_0)$.  The
conservation equation for the stress-energy tensor in spherical
symmetry is
\begin{equation}
{t_{AB}}^{|B}+2t_{AB}v^B=2Qv_A . \label{momconsbackground}
\end{equation}
As a manifestation of the contracted Bianchi identities,
(\ref{backgroundEinsteinaa}) can be obtained as a derivative of
equation (\ref{backgroundEinsteinAB}), provided that
(\ref{momconsbackground}) holds.

Now we introduce an arbitrary (not spherically symmetric) perturbation
of this spacetime: $g_{\mu\nu}\to g_{\mu\nu}(x^D)+h_{\mu\nu}(x^D,x^d)$
and again we perform a 2+2 decomposition. Furthermore we
decompose the angular ($x^d$) dependence into series of tensorial
spherical harmonics:  
\begin{itemize}
\item $Y^m_l(x^d)$ are the scalar spherical harmonics,
\item the objects ${Y^m_l}_{:a}$ and ${S^m_l}_a\equiv  \epsilon_a^{\ b}
{Y^m_l}_{:b}$ form a complete basis of vector harmonics, and
\item following Zerilli\cite{Zerilli}, we use the following basis of symmetric
tensor harmonics: $Y^m_l\gamma_{ab}$, ${Z^m_l}_{ab}\equiv
{Y^m_l}_{:ab}+\frac{l(l+1)}{2}Y^m_l\gamma_{ab}$, and
${S^m_l}_{a:b}+{S^m_l}_{b:a}$, which is a linear combination of the
basis introduced by Regge and Wheeler \cite{RW}.  For $l=0,1$ there is
only one linearly independent tensor, namely $\gamma_{ab}Y^m_l$,
while the other two tensors vanish. Gerlach and Sengupta initially
\cite{GS1}
used the Regge-Wheeler basis, but in \cite{GS2} changed to Zerilli's
basis in order to include the cases $l=0,1$ into a single formalism.
\end{itemize} 
All these spherical harmonics have definite parity under spatial
inversion: a spherical harmonic with label $l$ is called even if it
has parity $(-1)^l$ and odd if its parity is $(-1)^{l+1}$; $Y^m_l$,
${Y^m_l}_{:a}$ and ${Z^m_l}_{ab}$ are even and ${S^m_l}_a$ and
${S^m_l}_{a:b}+{S^m_l}_{b:a}$ are odd.  (An alternative terminology is
polar instead of even, and axial instead of odd.) Even and odd
perturbations decouple, and different values of $l$ and $m$
decouple. Furthermore, the perturbation equations do not depend on
$m$. In the following we consider one value of $l$ and $m$ at a time,
and suppress both the indices $l$ and $m$ and the explicit summation
over them. $h_{\mu\nu}$ is decomposed into
\begin{eqnarray}
h_{AB}&\equiv & {\tilde{h}_{AB}}Y , \\
h_{Ab}&\equiv & {h_A^{{\rm E}}} {Y}_{:b}+{h_A^{{\rm O}}} {S}_{b} , \\
h_{ab}&\equiv &r^2K \gamma_{ab}Y+ r^2G {Z}_{ab}+h ({S}_{a:b}+{S}_{b:a}) .
\end{eqnarray}
Note that the left-hand sides are components of a tensor on $M^4$.  On
the right-hand side $\tilde{h}_{AB}$ is a tensor on $M^2$, and $Y$ is
a scalar on $S^2$.  Similar remarks apply to the other definitions.
In the same way we decompose the perturbation $\Delta t_{\mu\nu}$ into
tensorial spherical harmonics:
\begin{eqnarray}
\Delta t_{AB}&\equiv & {\Delta \tilde{t}_{AB}}Y , \\
\Delta t_{Ab}&\equiv & {\Delta t_A^{{\rm E}}} {Y}_{:b}
+{\Delta t_A^{{\rm O}}} {S}_{b} , \\
\Delta t_{ab}&\equiv & r^2{\Delta t^3} \gamma_{ab}Y + {\Delta t^2} {Z}_{ab} 
                 +{\Delta t} ({S}_{a:b}+{S}_{b:a}) .
\end{eqnarray}
where we use the superindices $2$ and $3$ in order to follow the
notation of \cite{GS2}. (They are just labels, not components of 
any vector.)  Some
of the coefficients on the right hand side of these expansions are not
defined for $l=0,1$ because the corresponding spherical harmonics
vanish. In the following, we always point out which of the general
equations continue to hold for $l=0$ and $l=1$ if one sets these
coefficients to zero.

Now we define gauge-invariant variables, which do not contain
perturbations of the background generated by simple coordinate
transformations on this background.  With the shorthand
\begin{equation}
p_A\equiv h_A^{{\rm E}}-\frac{1}{2}r^2G_{|A} ,
\end{equation}
a complete set of gauge-invariant metric perturbations is
\begin{eqnarray}
k_{AB}&\equiv &\tilde{h}_{AB}-(p_{A|B}+p_{B|A}) ,\\
k_A&\equiv &h_A^{{\rm O}}-h_{|A}+2hv_A , \\
k&\equiv &K+\frac{l(l+1)}{2}G-2v^Ap_A .
\end{eqnarray}
A complete set of gauge-invariant matter perturbations is
\begin{eqnarray}
T_{AB}&\equiv &\Delta \tilde{t}_{AB}-t_{AB|C}p^C-t_{AC}p^C_{\ |B}
-t_{BC}p^C_{\ |A} , 
\\
T_A&\equiv &\Delta t_A^{{\rm E}}-t_{AC}p^C-\frac{r^2}{2}QG_{|A} , \\
T^3&\equiv &\Delta t^3-\frac{1}{r^2}(r^2 Q)_{|C}p^C+\frac{l(l+1)}{2}QG , \\
T^2&\equiv &\Delta t^2-r^2QG , \\
L_A&\equiv &\Delta t_A^{{\rm O}}-Qh_A , \\
L&\equiv &\Delta t-Qh .
\end{eqnarray}
(These are only partially gauge-invariant for $l=0,1$, and therefore
a partial gauge fixing is required in those cases in order to eliminate
coordinate transformations from the set of arbitrary perturbations.)

For $l\ge 2$ the metric and matter perturbations in any particular
gauge can be obtained by freely choosing values for $G$, $h^{{\rm
E}}_A$ and $h$ (the gauge in which these all vanish is
Regge-Wheeler gauge \cite{RW}) and then solving the definitions of the
gauge-invariant perturbations for the ``naked'' perturbations.

The perturbed Einstein equations, expressed only in gauge-invariant
perturbations, are
\begin{eqnarray}
l \ge 0: \hspace{14cm} \mbox{}  \nonumber \\
( k_{CA|B} + k_{CB|A} - k_{AB|C} ) v^C
- g_{AB} ( 2 {k_{CD}}^{|D} - {{k_D}^D}_{|C} ) v^C \nonumber
\\
- ( k_{|A} v_B + k_{|B} v_A + k_{|AB} )
+ \left( V_0 + \frac{l(l+1)}{2r^2} \right) k_{AB}
\nonumber \\
- g_{AB} \left[ {k^F}_F \frac{l(l+1)}{2r^2}
              + 2k_{DF} v^{D|F}
              + 3k_{DF} v^D v^F
              - {k_{|F}}^F
              - 3 k_{|F} v^F
              + \frac{(l-1)(l+2)}{2r^2} k
         \right]
&=& 8\pi  T_{AB}
,
\label{inveqAB}\\
\frac{1}{2} \{ - {k^{AB}}_{|AB}
               + {{{k^A}_A}^{|B}}_{|B}
               - 2 {k^{AB}}_{|A} v_B
               + {k^A}_{A|B} v^B
               + R^{AB} ( k_{AB} - k g_{AB} ) \nonumber \\
               - \frac{l(l+1)}{2r^2} k^A_{\ A}
               + {k_{|A}}^A
               + 2 k_{|A} v^A
            \}
&=& 8\pi  T^3
,
\label{inveq3}
\end{eqnarray}
\begin{eqnarray}
l \ge 1: \hspace{7cm} \mbox{}  \nonumber \\
\frac{1}{2} ( {k_{AB}}^{|B} - {k^B}_{B|A} + {k^B}_B v_A - k_{|A} )
&=& 8\pi  T_A
,
\label{inveqAeven}\\
- \frac{1}{2r^2} \left[ r^4 \left( \frac{k_A}{r^2} \right)_{|C}
                      - r^4 \left( \frac{k_C}{r^2} \right)_{|A}
                 \right]^{|C}
+ \frac{(l-1)(l+2)}{2r^2} k_A
&=& 8\pi  L_A
,
\label{inveqAodd}
\end{eqnarray}
\begin{eqnarray}
l \ge 2: \hspace{1cm} \mbox{}  \nonumber \\
- \frac{1}{2} {k^A}_A
&=& 8\pi  T^2
,
\label{inveq2}\\
\frac{1}{2} {k^A}_{|A}
&=& 8\pi  L
.
\label{inveqodd}
\end{eqnarray}
$R^{AB}$ in (\ref{inveq3}) are the $AB$ components of the
four-dimensional Ricci tensor.
Equations (\ref{inveq3}) and (\ref{inveqodd}) can be obtained as
derivatives of the other equations using the linearized equations of
stress-energy conservation, which are
\begin{eqnarray}
l \ge 0: \hspace{6cm} \mbox{}  \nonumber \\
\frac{1}{r^2} ( r^2 T_{AB} )^{|B}
- \frac{l(l+1)}{r^2} T_A - 2 v_A T^3
&=&
( t_{AB|D} + 2 t_{AB} v_D ) k^{BD}
+ Q ( k_{|A} -2 k v_A )
- t_{AB} k^{|B}
\nonumber \\
&&
+ \frac{1}{2} t^{BC} k_{BC|A}
- \frac{1}{2} t_{AB} {k_F}^{F|B}
+ t_{AB} {k^{BF}}_{|F}
,
\label{momconsA}
\end{eqnarray}
\begin{eqnarray}
l \ge 1: \hspace{6cm} \mbox{}  \nonumber \\
\frac{1}{r^2} ( r^2 T_B )^{|B}
+ T^3
- \frac{(l-1)(l+2)}{2} \frac{T^2}{r^2}
&=&
\frac{1}{2} k_{AB} t^{AB}
+ Q \left( k - \frac{1}{2} {k^A}_A \right)
,
\label{momconseven}\\
( r^2 L_B )^{|B}
&=&
(l-1)(l+2) L
.
\label{momconsodd}
\end{eqnarray}


\section{Perturbation equations for arbitrary matter 
in an arbitrary orthonormal basis}


Both in order to transform tensor equations into sets of scalar
equations, and in order to separate evolution equations from
constraints, it is desirable to introduce an orthonormal frame
in $M^2$, namely
\begin{equation}
-u_Au^A\equiv n_An^A\equiv 1 , \ \ \ \ u_A n^A\equiv 0 .
\end{equation}
In the presence of curvature, this cannot be a coordinate basis:
\begin{equation}
[n,u]^A=\mu n^A-\nu u^A  , \ \ \ \ 
\mu\equiv {u^A}_{|A} , \ \ \ \ \nu\equiv {n^A}_{|A}.
\end{equation}
We define an associated basis of 2-tensors,
\begin{eqnarray}
g_{AB}=-u_A u_B+n_A n_B , \ \ \ \ \epsilon_{AB}\equiv n_A u_B-u_A n_B , \\
p_{AB}\equiv  u_A u_B+n_A n_B , \ \ \ \ q_{AB}\equiv n_A u_B+u_A n_B .
\end{eqnarray}
and use it to decompose the gauge-invariant metric perturbation:
\begin{equation}
k_{AB}\equiv \eta g_{AB} + \phi p_{AB} + \psi q_{AB} .
\end{equation}
We define derivatives along the basis vectors:
\begin{equation}
\dot f\equiv u^A f_{|A} , \ \ \ \ f^\prime\equiv n^A f_{|A} ,
\end{equation}
and re-express the even-perturbation equations in this basis. We also
introduce the
notation 
\begin{equation}
u^Av_A\equiv U, \quad n^Av_A\equiv W, \quad W^2-U^2=v^A v_A\equiv
v^2.
\end{equation}
For reference we give the background Einstein equations in frame components:
\begin{eqnarray}
W'-\dot U+\nu W-\mu U-2U^2+2W^2-r^{-2}&=&8\pi \frac{1}{2}{t_A}^A \ , 
\label{Einsteineqbasis1}\\
-W'-\dot U+\nu W+\mu U-U^2-W^2&=&8\pi \frac{1}{2}p_{AB}t^{AB}  \ , 
\label{Einsteineqbasis2}\\
-U'-\dot W+\mu W+\nu U-2UW&=&8\pi \frac{1}{2}q_{AB}t^{AB}  \ , 
\label{Einsteineqbasis3}\\
-{\cal R}+W^\prime-\dot U+\nu W-\mu U-U^2+W^2&=&8\pi Q \ . 
\label{Einsteineqbasis4}
\end{eqnarray}
We use them among other things to bring all perturbation equations
into a standard form by eliminating the derivatives of $U$ and $W$.

The complete Einstein equations for the even perturbations, still for
arbitrary matter content, expressed in gauge-invariant variables, and
decomposed in an arbitrary frame, are
\begin{eqnarray}
l \ge 0 : \hspace{13cm} \mbox{} && \nonumber \\
- \ddot{k} + k'' + \nu k' - \mu \dot{k}
- 2 U [ 2 \dot{k} + \dot{\phi} + \psi' + 2 \mu \phi + 2 \nu \psi ]
- 2 W [ - 2 k' + \phi' + \dot{\psi} + 2 \nu \phi + 2 \mu \psi ]  
&& \nonumber \\
- \frac{l(l+1)}{r^2} ( k + \eta ) + \frac{2}{r^2} ( k - \eta )
- 2 \phi ( U^2 + W^2 ) - 4 \psi UW 
+ 16 \pi  ( \phi p_{AB} + \psi q_{AB} ) t^{AB}
&=& 8\pi  {T_A}^A 
,
\label{inveqbasis1}\\
- \ddot k - k'' + \nu k' + \mu \dot k
+ 2 U [ - \dot k + \dot\eta + \psi' + 2 \mu \phi ]
+ 2 W [ - k' + \eta' - \dot{\psi} - 2 \nu \phi ] 
&& \nonumber \\
+ 2 \phi \left( V_0 + \frac{l(l+1)}{2r^2} \right)
&=& 8 \pi  p^{AB} T_{AB}
,
\label{inveqbasis2}\\
- ( \dot k )' - ( k' ) \dot{} + \mu k' + \nu \dot k
+ 2 U [ - k' + \eta' - \phi' - 2 \mu \psi ]
+ 2 W [ - \dot k + \dot\eta + \dot{\phi} + 2 \nu \psi ]
&& \nonumber \\
- 2 \psi \left( V_0 + \frac{l(l+1)}{2r^2} \right)
&=& 8 \pi  q^{AB} T_{AB} 
,
\label{inveqbasis3}\\
-\ddot k -\ddot \eta - \ddot\phi + k'' + \eta''- \phi'' 
- (\dot\psi)' - (\psi')\dot{}
+ \nu k' - \mu\dot{k}
+ \nu\eta' - \mu\dot\eta
- 3 ( \nu\phi' + \nu\dot\psi + \mu\psi' + \mu\dot\phi )
&& \nonumber \\
- 2 {U} ( \dot\phi + \dot{k} + \psi' )
- 2 {W} ( \phi' - k' + \dot\psi )
- 2 ( \nu'\phi + \dot\nu\psi + \mu'\psi + \dot\mu\phi )
- \frac{l(l+1)}{r^2} \eta
&& \nonumber \\
- 2 ( k - \eta )
    [ {\cal R} - {W}' + \dot{U} - \nu{W} + \mu{U} + {U}^2 - {W}^2
]
&& \nonumber \\
- 2 \phi
    [ {W}' + \dot{U} + \nu{W} + \mu{U} + {U}^2 + {W}^2 + \mu^2 + \nu^2 ]
&& \nonumber \\
- 2 \psi
    [ {U}' + \dot{W} + \mu{W} + \nu{U} + 2{U}{W} + 2\mu\nu ]
&=&
16 \pi  T^3, 
\label{inveqbasis4} \\
l \ge 1 : \qquad \dot k + \dot\eta + \dot{\phi} + \psi' + 2 \mu \phi 
+ 2 \nu \psi - 2 U \eta &=& - 16 \pi  u^A T_A 
,
\label{inveqbasis5}\\
k' + \eta' - \phi' - \dot{\psi} - 2 \nu \phi - 2 \mu \psi - 2 W \eta 
&=& 
- 16 \pi  n^A T_A 
,
\label{inveqbasis6}\\
l \ge 2 : \qquad \eta &=& - 8 \pi  T^2 
. 
\label{inveqbasis7}
\end{eqnarray}

We have now turned the even perturbation equations into scalar
form. They are already in first-order form if one counts $f$, $\dot f$
and $f'$ as separate variables linked by certain trivial equations. In
the following we always imply this first-order interpretation. The
final step of the analysis is to separate the equations into evolution
equations and constraints. This cannot be done in general, as the
causal structure of the equations depends on the matter content.

Attention must be paid to the regularity of the perturbations at
$r=0$.  Changing to Cartesian coordinates one can see that regular
perturbations scale as
\begin{eqnarray}
\eta &=& r^l \bar\eta , \label{regularity_eta} \\
k &=& r^l \bar k , \label{regularity_k} \\
\psi &=& r^{l+1} \bar\psi , \label{regularity_psi}\\
\chi \equiv  \phi - k + \eta &=& r^{l+2} \bar\chi , \label{regularity_chi}
\end{eqnarray}
where the barred variables are $O(1)$ at the center. 

The odd metric-perturbations are contained in $k_A$. We can transform
the vector equation (\ref{inveqAodd}) into a scalar equation using the curl 
of $k_A$:
\begin{equation}
\Pi\equiv \epsilon^{AB}(r^{-2}k_A)_{|B} .
\end{equation}
It is possible to reconstruct $k_A$ from $\Pi$ for $l\ge 2$ using
equation (\ref{inveqAodd}). Therefore
$\Pi$ alone characterizes the physical odd metric perturbations. 
For $l\ge 2$ it
obeys the ``odd parity master equation'' \cite{GS1}
\begin{equation}
l\ge 2 : \qquad
-\frac{1}{2}\left[\frac{1}{r^2}(r^4\Pi)^{|A}\right]_{|A}
+\frac{(l-1)(l+2)}{2}\Pi=8\pi  \epsilon^{AB}L_{A|B} . \label{Oddeq}
\end{equation}
This equation is a generalization of the Regge-Wheeler equation\cite{RW}.
If we define the object $\Pi_{RW}=r^3\Pi$ then the master equation is:
\begin{equation}
{{\Pi_{RW}}^{|A}}_A
+\left({v^A}_{|A}-v^Av_A-\frac{(l-1)(l+2)}{r^2}\right)\Pi_{RW}
=
-16\pi r \epsilon^{AB}L_{A|B}
\end{equation}
which, for Schwarzschild background in radial coordinates, and using the 
``tortoise'' coordinate $r^*$, is the Regge-Wheeler equation:
\begin{equation}
-\frac{\partial^2\Pi_{RW}}{\partial t^2}
+\frac{\partial^2\Pi_{RW}}{\partial r^{*2}}
+\left(1-\frac{2M}{r}\right)\left(\frac{6M}{r^3}
-\frac{l(l+1)}{r^2}\right)\Pi_{RW}=0.
\end{equation}

We enforce regularity at the origin by defining
\begin{equation}
\Pi=r^{l-2}\bar\Pi ,
\end{equation}
and equation (\ref{Oddeq}) in an arbitrary basis becomes
\begin{eqnarray}
l \ge 2: \qquad -\ddot{\bar\Pi}
+{\bar\Pi}^{\prime\prime}
+[\nu+2(l+1)W]\bar\Pi^\prime-[\mu+2(l+1)U]\dot{\bar\Pi} && \nonumber \\
+(l+2)\left[\frac{{r_{|A}}^A}{r}+(l-1)(v_Av^A-r^{-2})\right]\bar\Pi&=&-16\pi 
 r^{-l}[-(u_AL^A)^\prime+(n_AL^A)\dot{}+\mu n_AL^A-\nu u_A L^A] . 
\label{Oddbareqbasis}
\end{eqnarray}

In the special case $l=1$, $k_A$ is defined by $\Pi$ only up to a
gradient, but precisely this gradient is a gauge degree of freedom, so
that $\Pi$ again contains all the gauge-invariant information. As we
have $(r^2L_A)^{|A}=0$, $L_A$ can be expressed as
$r^2L_A=\epsilon_{AB}T^{|B}$, with $T$ a new scalar. Equation
(\ref{inveqAodd}) can be integrated to obtain the algebraic Einstein equation
\begin{equation}
l=1: \qquad r^4\Pi=16\pi  T + {\rm const}. \label{Oddeql=1}
\end{equation}
The integration constant must be zero if the perturbed spacetime
is to be regular in $r=0$. (If the background spacetime is
Schwarzschild, then this integration constant parameterizes an
infinitesimal angular momentum taking Schwarzschild into Kerr.)
For $l=0$, there are no odd-parity perturbations at all.


\section{The massless scalar field model}


In the remainder of the paper, we restrict attention to a particular
matter model, the real massless scalar field $\varphi$ with
stress-energy tensor
\begin{equation}
t_{\mu\nu}
=
\varphi_{,\mu}\varphi_{,\nu}
-\frac{1}{2}g_{\mu\nu}\varphi_{,\lambda}\varphi^{,\lambda}
.
\end{equation}
The background momentum-conservation equation
(\ref{momconsbackground}) gives the evolution equation of the field:
\begin{equation}
\frac{1}{r^2}(r^2\varphi_{|A})^{|A}=\varphi_{|A}^{\ \ A}+2v^A\varphi_{|A}=0 .
\label{backgroundfieldeq}
\end{equation}
It is useful to notice that for  scalar field matter
\begin{equation}
V_0=\frac{{r_{|A}}^{|A}}{r}=r^{-2}-v_Av^A = {2m_{\rm Hawking}\over
r^3}.
\end{equation}

The scalar field has a perturbation $\sum_{l,m}Y\Delta\varphi$. We can
construct a gauge-invariant perturbation as
\begin{equation}
\Phi\equiv \Delta\varphi-p^C\varphi_{|C} ,
\end{equation}
in terms of which the gauge-invariant perturbations of the
stress-energy tensor are
\begin{eqnarray}
T_{AB}&=&\Phi_{|A}\varphi_{|B}
+\Phi_{|B}\varphi_{|A}-g_{AB}\varphi_{|F}\Phi^{|F}
+\frac{1}{2}g_{AB}k^{DF}\varphi_{|D}\varphi_{|F}
-\frac{1}{2}k_{AB}\varphi_{|F}\varphi^{|F} , \label{invmatterAB}\\
T_A&=&\Phi\varphi_{|A} , \label{invmatterA}\\
T^3&=&kQ-\varphi_{|D}\Phi^{|D}+\frac{1}{2}k^{DF}\varphi_{|D}\varphi_{|F} , \\
T^2&=&0 , \\
L_A&=&0 , \\
L&=&0 .
\end{eqnarray}
Notice that there are no odd perturbations.

Again, the momentum-conservation equation (\ref{momconsA}) gives the evolution
equation for the matter perturbation, that is, the perturbed scalar wave 
equation:
\begin{equation}
l\ge 0: \qquad \frac{1}{r^2}(r^2\Phi_{|A})^{|A}
-\frac{l(l+1)}{r^2}\Phi=\frac{1}{r^2}(r^2k_{AB}\varphi^{|A})^{|B}
-(k+\eta)_{|B}\varphi^{|B}
.
\label{inveqwave}
\end{equation}
Equations (\ref{momconseven}) and (\ref{momconsodd}) are redundant for
scalar field matter.
If $\varphi=0$, matter and metric perturbations decouple.

To enforce regularity at the origin, we define
\begin{equation}
\Phi=r^l\bar\Phi , \label{regularity_Phi}
\end{equation}
where $\bar\Phi$ is $O(1)$ at $r=0$.


\section{Choice of frame and coordinate system}


In the remainder of the paper we shall use the radial basis defined by
\begin{equation}
n^A\equiv \frac{v^A}{v} 
\qquad W=v , \qquad U=0 \label{vbasis} .
\end{equation}
There is a system of coordinates naturally associated with this basis, which
uses $r$ as a coordinate: the familiar ``Schwarzschild-like'' coordinate 
system, in which the metric is
\begin{equation}
ds^2 = -\alpha^2(t,r) dt^2 + a^2(t,r) dr^2 + r^2 d\Omega^2,
\label{Schwarzschild-like_metric}
\end{equation}
In these coordinates, the derivatives in the $v_A$ frame take the form
\begin{equation}
\dot f = \alpha^{-1} {\partial f \over \partial t} , \qquad
f'  =  a^{-1} {\partial f \over \partial r} .
\end{equation}
This is not yet the coordinate system we shall use, but it is useful
as an intermediate step in the presentation of the final coordinates.

The Choptuik critical solution is a solution of the Einstein-real
massless scalar field system defined by its self-similarity together
with regularity. We introduce coordinates $x$ and $\tau$ adapted to
self-similarity of the spacetime. The background solution has the
geometric property of being discretely self-similar (DSS), which in
our coordinates means that
$\varphi(\tau+\Delta,x)=\varphi(\tau,x)$. The metric coefficients $a$
and $\alpha$ defined in (\ref{Schwarzschild-like_metric}) have the same
periodicity. Coordinates with this property are not unique.  We make the
following choice (in terms of the Schwarzschild-like coordinates):
\begin{equation}
\tau \equiv  - \ln\left(-\frac{t}{r_0}\right) , \qquad 
x \equiv  \left(-{r\over t}\right) e^{-\xi_0(\tau)} ,
\end{equation}
where $r_0$ is an arbitrary scale. In the following we set it equal
to 1.  Our choice has the following properties.  Surfaces of constant
$\tau$ coincide with those of constant $t$, and $\tau$ increases with
$t$. Therefore $\tau$ is a good time coordinate, as well as being the
logarithm of overall spacetime scale. The origin $r=0$ coincides with
$x=0$.  We choose the function $\xi_0(\tau)$ such that the past light
cone of the point $r=t=0$ coincides with the surface $x=1$.  The
domain of dependence of the disk $0\le x\le1$ on any spacelike surface
is therefore given by $0\le x\le1$. We can therefore work on the
numerical domain $0\le x\le1$, $0\le\tau<\infty$ without requiring
boundary data on $x=1$. If we extended our perturbation initial data
to $x>1$, that part of the data could not influence $x<1$. We can
therefore determine exponential growth or decay on the domain
$0\le x\le1$ alone.

In these coordinates the frame derivatives in the radial frame are
\begin{eqnarray}
\dot f & = & \alpha^{-1} e^\tau \left[{\partial f \over \partial \tau} +
\left(1-{d\xi_0\over d\tau}\right) x {\partial f \over \partial x} \right] , \\
f' & = & a^{-1}e^{\tau-\xi_0} {\partial f \over \partial x} ,
\end{eqnarray}
and the spacetime metric in these coordinates, but expressed through
$a$ and $\alpha$, is 
\begin{equation}
ds^2 =
r_0^2 e^{-2\tau}\left\{
-\alpha^2 d\tau^2 
+ a^2e^{2\xi_0} \left[dx 
+ \left(1-{d\xi_0\over d\tau}\right) x \, d\tau\right]^2 +
x^2 e^{2\xi_0}\,d\Omega^2\right\} .
\end{equation}

The background Einstein equations and a few more definitions are given
in appendix \ref{appendixA}.


\section{Odd perturbations of the Choptuik spacetime}


As we have seen, both $L_A$ and $L$ vanish and therefore the odd
metric perturbations decouple from the matter perturbations. This
implies [see equation (\ref{Oddeql=1})] that for $l=1$ we have $T=0$,
and hence $\Pi=0$, if we demand regularity at the center. $k_A$ is
then pure gauge. All $l=1$ odd perturbations are therefore pure
gauge. For $l\ge 2$ equation (\ref{Oddbareqbasis}) is, in the radial
basis,
\begin{equation}
l \ge 2: \qquad -\ddot{\bar\Pi}
+{\bar\Pi}^{\prime\prime}+[\nu+2(l+1)v]\bar\Pi^\prime-\mu\dot{\bar\Pi}
- (l^2-4)V_0\bar\Pi= 0 .
\end{equation}

This equation is equivalent to the first-order system
\begin{eqnarray}
-(\dot{\bar\Pi})\dot{} + ({\bar\Pi}')' + S_1 & = & 0 , \\
-({\bar\Pi}')\dot{} + (\dot{\bar\Pi})' + S_2 & = & 0 , \label{firstid}\\
-({\bar\Pi})\dot{}  + \dot{\bar\Pi} & = & 0 , \\
-({\bar\Pi})' + {\bar\Pi}' & = & 0 , \label{lastid}
\end{eqnarray}
where 
\begin{eqnarray}
S_1 & = & - \mu \dot{\bar\Pi} + [\nu+2(l+1)v] {\bar\Pi}' - (l^2-4)V_0 
{\bar\Pi} 
, \\
S_2 & = & -\mu {\bar\Pi}' + \nu \dot{\bar\Pi} .
\end{eqnarray}
Note that equations (\ref{firstid}-\ref{lastid}) are really identities
that need to be added to the system when we consider $\bar\Pi$,
$\dot{\bar\Pi}$ and $\bar\Pi'$ as independent variables. From this
first-order point of view, we now have three evolution equations
(which contain dot-derivatives) and one constraint (which does
not). The three characteristics are the light rays and the lines of
constant $r$. Note that this causal structure is independent of any
particular choice of coordinates. Now we introduce coordinates
$(\tau,x)$.  We also rescale ${\bar\Pi}$ and its derivatives so that
the rescaled variables $u$ are precisely periodic in $\tau$ if (and
only if) the perturbed solution is DSS.  Consider a perturbation of a
self-similar background so that the sum of background and
perturbations is again self-similar (to linear order in the
perturbations). To find the scaling behavior of $\Pi$, we note that
the tensor $k_A S_a$ must scale like the metric itself. $S_a$ scales
trivially, so that $k_A$ scales like the metric itself. On the other
hand $\epsilon^{AB}$ scales like the inverse metric, so that the
scalar $\epsilon^{AB}k_{A|B}$ scales trivially, that is, it is
periodic in $\tau$ for a DSS perturbation. Therefore $\bar\Pi =
r^{2-l}
\epsilon^{AB}(r^{-2}k_A)_{|B}$ scales like $r^{-l}$, that is like
$e^{l\tau}$. We also note that each frame derivative adds a power
$e^\tau$. In order to cancel this scaling behavior, we define
\begin{eqnarray}
u_1 & \equiv  & e^{-(l+1)\tau} \dot{\bar\Pi} , \\
u_2 & \equiv  & e^{-(l+1)\tau} {\bar\Pi}' , \\
u_3 & \equiv  & e^{-l\tau} {\bar\Pi} .
\end{eqnarray}
The final form of the equations is then
\begin{eqnarray}
{\partial u\over \partial \tau} + A_3 {\partial u\over \partial x} + s = 0 , 
\end{eqnarray}
where the 3$\times$3 matrix $A_3$ is
\begin{equation}
A_3 \equiv  {\rm diag}(A_2, \lambda_0)
\qquad \lambda_0\equiv  x \left(1-{d\xi_0\over d\tau}\right) .
\end{equation}
with
\begin{equation} 
A_2 \equiv  \left(\matrix{
\lambda_0 & -{(\alpha / a)} e^{-\xi_0}  \cr 
-{(\alpha / a)} e^{-\xi_0} & \lambda_0  \cr
}\right) .
\label{oddAmatrix}
\end{equation}

We first consider the transport part of the equations.  The
characteristic speeds, or eigenvalues of $A_3$ are
\begin{equation}
\lambda_0, \quad \lambda_\pm = \lambda_0 \pm {\alpha \over a}
e^{-\xi_0} .
\end{equation}
$\lambda_0$ and $\lambda_+$ are always positive, while $\xi_0$ has
been chosen so that $\lambda_-$
changes sign at $x=1$ by definition. That is, $\xi_0(\tau)$ is defined
by the equation
\begin{equation}
\left(1-{d\xi_0\over d\tau}\right) \, e^{\xi_0} \equiv  \left({\alpha \over
a}\right)_{x=1} .
\end{equation}
This definition means that for $0\le x < 1$ the characteristic speeds
$\lambda_0$ and $\lambda_+$ are positive, and $\lambda_-$ is negative.
At $x=1$, $\lambda_0$ and $\lambda_+$ are still positive, and
$\lambda_-$ is zero. Therefore no boundary condition is required at
the boundary $x=1$, because no information
crosses it from the right. At $x=0$, all $u$ are either even or odd in
$x$, so that boundary conditions are obtained trivially. 

The source terms in the final equations are
\begin{eqnarray}
s_1 & = & - \alpha e^{-(l+2)\tau} S_1 + (l+1) u_1
= - \alpha\left[-\bar \mu u_1 + \bar\nu u_2 + 2(l+1)\bar v u_2 
- (l^2-4) \bar V_0 u_3 \right] + (l+1) u_1 , \\
s_2 & = &  - \alpha e^{-(l+2)\tau} S_2 + (l+1) u_2
= - \alpha\left[-\bar \mu u_2 + \bar\nu u_1 \right] + (l+1) u_2 , \\
s_3 & = &  - \alpha u_1 + l u_3 .
\end{eqnarray}
We have used rescaled background coefficients that are periodic in
$\tau$ on the DSS background. Using the background Einstein equations
they are
\begin{eqnarray}
\bar\mu & \equiv  & e^{-\tau} \mu = {2a \over e^{\xi_0} x} XY , \\
\bar\nu & \equiv  & e^{-\tau} \nu= {a \over e^{\xi_0} x} \left(X^2+Y^2
+ {1-a^{-2} \over2}\right) , \\
\bar v & \equiv  & e^{-\tau} v = {1\over e^{\xi_0} a x} , \\
\bar V_0 & \equiv  & e^{-2\tau} V_0 = {1-a^{-2} \over e^{2\xi_0} x^2} . 
\end{eqnarray}
Note that at $x=0$, these are regular except for $\bar v \sim x^{-1}$.
(The background fields $X$ and $Y$ are defined in the appendix.)
Finally, the constraint equation becomes
\begin{equation}
{\partial u_3\over \partial x} = a e^{\xi_0} u_2 .
\end{equation}
As free initial data we can take $u_1$ and $u_3$, and we obtain $u_2$
by taking the derivative of $u_3$. Numerically it is more stable to
take $u_1$ and $u_2$, plus the value of $u_3$ at $x=0$, as free data,
and solve for $u_3$ by integration. 


\section{Even perturbations of the Choptuik spacetime}


\subsection{General case $l\ge 2$}


The even perturbation equations are far more complicated. 
We discuss the cases $l\ge2$, $l=0$ and $l=1$ separately, beginning
with the general case $l\ge2$. 

The vanishing of the matter perturbation $T^2$ makes $k_{AB}$ traceless
($\eta=0$). Therefore the even perturbations are described by
$\bar\chi,\bar\psi,\bar k$ and $\bar\Phi$. These obey the following
set of equations:
\begin{eqnarray}
- \ddot{\bar\Phi} + \bar{\Phi}'' - \mu\dot{\bar\Phi}
+ ( \nu + 2(l+1)v ) \bar{\Phi}'
+ \left( - l^2 V_0 + \frac{8}{r^2} ( Y^2 - X^2 ) \right) \bar\Phi
\nonumber \\
- 2 ( \bar k + r^2 \bar\chi ) \frac{1}{r} ( \dot Y - \nu X )
- 2 \bar\psi ( \dot X + ( v - \nu ) Y )
&=& 0 , \\
- \ddot{\bar k} + \bar{k}'' - \mu\dot{\bar k}
+ ( \nu + 2 (l+1) v ) \bar{k}' - l^2 V_0 \bar k - 2\bar\chi
+ 2 \left( V_0 + \frac{2}{r^2} ( X^2 + Y^2 ) \right) ( \bar k + r^2 \bar\chi )
+ \frac{8}{r} X Y \bar\psi
- \frac{16v}{r} X \bar\Phi
&=& 0  , \\
r^2 \dot{\bar\chi} + 2 \dot{\bar k} + r \bar{\psi}' 
+ ( 2 \nu + (l+1) v ) r \bar\psi + 2 \mu ( \bar k + r^2 \bar\chi )
+ \frac{8}{r} Y \bar\Phi 
&=& 0  , \\
r \dot{\bar\psi} + r^2 \bar{\chi}' + (l+2) r^2 v \bar\chi
+ 2 \nu ( \bar k + r^2 \bar\chi ) + 2 \mu r \bar\psi - \frac{8}{r} X \bar\Phi
&=& 0 , \\
\bar{k}'' + 2 (l+1) v \bar{k}' - \mu \dot{\bar k}
- \left( \frac{(l+1)(l+2)}{2} + 2 r^2 v^2 \right) \bar\chi
- ( l(l+1) V_0 + l \nu v ) \bar k - v r^2 \bar{\chi}' 
\nonumber \\
+ \left( l V_0 + \frac{4}{r^2} Y^2 \right)( \bar k + r^2 \bar\chi )
+ \frac{4}{r} ( Y \dot{\bar\Phi} + X ( l v \bar\Phi + \bar{\Phi}' ) )
+ r \bar\psi \left( - 2 \mu v + \frac{4}{r^2} X Y \right)
&=& 0  , \\
- ( \dot{\bar k} )' - (l+2) v \dot{\bar k} + \mu \bar{k}' + (l-2) \mu v \bar k
- r v \bar{\psi}' 
- r \bar\psi 
    \left( (l+1) v^2 + V_0 + \frac{l(l+1)}{2r^2} + \frac{2}{r^2} ( X^2 - Y^2 ) 
    \right)
- 2 \mu v r^2 \bar\chi 
\nonumber \\
- \frac{4}{r} ( X \dot{\bar\Phi} + Y \bar{\Phi}' )
- \frac{4}{r} v Y \bar\Phi (l+2)
&=& 0 .
\end{eqnarray}
Again, we rescale the variables so that they are periodic in $\tau$ if
and only if the perturbed spacetime is still DSS:
\begin{eqnarray}
    u_1&\equiv &e^{-(l+1)\tau}\dot{\bar\Phi} ,
    u_2\equiv e^{-(l+1)\tau}\bar{\Phi}^\prime ,
    u_3\equiv e^{-l\tau}\bar{\Phi} , \nonumber \\
    u_4&\equiv &e^{-(l+1)\tau}\dot{\bar k} ,
    u_5\equiv e^{-(l+1)\tau}\bar{k}^\prime ,
    u_6\equiv e^{-l\tau}\bar{k} , \nonumber \\
    u_7&\equiv &e^{-(l+1)\tau}r\bar{\chi} ,
    u_8\equiv -e^{-(l+1)\tau}\bar{\psi} .
\label{firstordervariables}
\end{eqnarray}
There are 8 evolution equations of the form
\begin{equation}
\frac{\partial u}{\partial\tau}+A_8\frac{\partial u}{\partial x}+s=0 ,
\end{equation}
where the matrix $A_8$ is 
\begin{equation}
A_8 \equiv  {\rm diag}\left(A_3, A_3, A_2\right),
\end{equation}
and $s$ is the vector
\begin{eqnarray}
s_1 & = &
      (l+1) u_1
      - \alpha
            \left[
            - \bar\mu u_1
            + ( \bar\nu + 2(l+1) \bar v ) u_2
            + \left( -l^2 \bar{V}_0
                     + \frac{8}{x^2 e^{2\xi_0}} (Y^2-X^2)
              \right) u_3
            \right.
\nonumber \\ &&
            \ \ \ \ \ \ \ \ \ \ \ \ \ \ \ \ \ \ \
            \left.
            - 2 (e^{-\tau}\dot Y - \bar\nu X) 
                \left( \frac{1}{x e^{\xi_0}} u_6 + u_7 \right)
            + 2 (e^{-\tau}\dot X + (\bar v-\bar\nu) Y) u_8
            \right]
, \\
s_2 & = & 
      (l+1) u_2
      - \alpha [ \bar\nu u_1 - \bar\mu u_2 ]
, \\
s_3 & = & l u_3 - \alpha u_1
,  \\
s_4 & = & 
      (l+1) u_4
      - \alpha
            \left[
            - \frac{16}{x e^{\xi_0}} \bar v X u_3
            - \bar\mu u_4
            + ( \bar\nu + 2(l+1) \bar v )u_5
            + \left( (2-l^2) \bar{V}_0
                    + \frac{4}{x^2 e^{2\xi_0}} (Y^2+X^2)
              \right) u_6
            \right.
\nonumber \\ &&
            \ \ \ \ \ \ \ \ \ \ \ \ \ \ \ \ \ \ \
            \left.
            + \frac{2}{x e^{\xi_0}} (-a^{-2} + 2 (X^2+Y^2) ) u_7
            - \frac{4}{a} \bar\mu u_8
            \right]
, \\
s_5 & = &
      (l+1) u_5
      - \alpha [ \bar\nu u_4 - \bar\mu u_5 ]
, \\
s_6 & = & l u_6 - \alpha u_4
,  \\
s_7 & = &
      (l+1) u_7
      - \alpha
            \left[
            - \frac{8}{x^2 e^{2\xi_0}} Y u_3
            - \frac{2}{x e^{\xi_0}} ( u_4 + \bar\mu u_6 )
            - 2 \bar\mu u_7
            + ( 2 \bar\nu + (l+1) \bar v) u_8
            \right]
, \\
s_8 & = &
      (l+1) u_8
      - \alpha
            \left[
            - \frac{8}{x^2 e^{2\xi_0}} X u_3
            + \frac{2}{x e^{\xi_0}} \bar\nu u_6
            + ( 2 \bar\nu + (l+1) \bar v) u_7
            - 2 \bar\mu u_8
            \right]
.
\end{eqnarray}

There are four constraints
\begin{eqnarray}
\frac{\partial u_3}{\partial x} & = & a e^{\xi_0} u_2
, \\
\frac{\partial u_6}{\partial x} & = & a e^{\xi_0} u_5
, \\
\frac{\partial u_7}{\partial x} + \frac{b_7}{x} u_7 & = & c_7
, \\
\frac{\partial u_8}{\partial x} + \frac{b_8}{x} u_8 & = & c_8
,
\end{eqnarray}
where
\begin{eqnarray}
b_7 & = & a^2 \left( \frac{2+l+l^2}{2}
         + \frac{l+1}{a^2}
         - 4 Y^2
        \right)  , \\
b_8 & = & a^2  \left( \frac{2+l+l^2}{2}
         + \frac{l}{a^2}
         + 2 (X^2-Y^2)
        \right)  , \\
c_7 & = & a \frac{\partial u_5}{\partial x}  
      - a^2 e^{\xi_0}  \left\{
	- \frac{4}{x e^{\xi_0}} ( Y u_1 + X ( u_2 + l \bar v u_3 ) )
      + \bar\mu u_4
      -2(l+1) \bar v u_5
      + \left( l^2 \bar{V}_0
         + l \bar\nu \bar v
         - \frac{4}{x^2 e^{2\xi_0}} Y^2
        \right) u_6 \right\}
, \\
c_8 & = &  a \frac{\partial u_4}{\partial x}
	- a^2 e^{\xi_0} \left\{
      - \frac{4}{x e^{\xi_0}} ( X u_1 + Y ( u_2 + (l+2) \bar v u_3 ) )
      -(l+2) \bar v u_4
      + \bar\mu u_5
      + (l-2) \bar\mu \bar v u_6
      - \frac{2}{a} \bar\mu u_7
      \right\} .
\end{eqnarray}
The causal structure of the equations is similar to the odd case,
because $A_8$ is constructed from $A_2$ and $A_3$. The characteristics
of $A_2$ are just the ingoing and outgoing radial null
geodesics. $u_1$, $u_2$ and $u_3$ on the one hand, and $u_4$, $u_5$
and $u_6$ each form a wave equation with a mass-like term, while $u_7$
and $u_8$ form a massless wave equation. The first two constraints are
also identical to the odd perturbation case, and can be solved for
$u_3$ and $u_6$ by integration, or for $u_2$ and $u_5$ by
differentiation. Again we choose the former in the numerical
treatment, taking the value of $u_3$ and $u_6$ at $x=0$ as free
initial data, together with $u_1$, $u_2$, $u_4$ and $u_5$.

The next constraint equation contains $u_7$ but not $u_8$, and
is therefore a linear ordinary differential equation (ODE)
 for $u_7$. Once $u_7$ is known, the
last constraint can be solved as an ODE for $u_8$. We solve these ODEs
by a second-order implicit method, in order to finite-difference all
constraints in the same way.  Both the evolution equation for $u_7$
and the constraint for $u_8$ require the following condition at the
origin $x=0$ for all $\tau$ in order to be consistent:
\begin{equation}
2 u_4 + {8\over e^{\xi_0}} {\partial Y\over \partial 
x}u_3 - (l+1) u_8 = O(x^2) .
\end{equation}
We solve this constraint for the value of $u_8$ at $x=0$. The value of
$u_7$ at $x=0$ is zero by definition. These boundary conditions
complete the constraints for $u_7$ and $u_8$, which are then
determined completely, given $u_1$ to $u_6$.


\subsection{Special case $l=0$}


For $l=0$ a general perturbation is described by the objects
($k_{AB},k,T_{AB},T^3$), which are not gauge-invariant: under an
arbitrary coordinate transformation generated by the vector $\xi_\mu
dx^\mu=\tilde{\xi}_A Y dx^A$ these objects change as
\begin{eqnarray}
k_{AB} &\longrightarrow& k_{AB}-(\tilde{\xi}_{A|B}+\tilde{\xi}_{B|A}) , \\
k &\longrightarrow& k - 2v^A \tilde{\xi}_A , \\
T_{AB} &\longrightarrow& T_{AB} - t_{AB|C}\tilde{\xi}^C
-t_{AC}{\tilde{\xi}^C}_{\ \ |B}-t_{BC}{\tilde{\xi}^C}_{\ \ |A} , \\
T^3 &\longrightarrow& T^3 -\frac{1}{r^2}(r^2Q)_{|D}\tilde{\xi}^D .
\end{eqnarray}
Therefore we have to impose two gauge conditions. In our case we want
to maintain the form (\ref{Schwarzschild-like_metric}) of the metric
during perturbation, so we perform a gauge transformation to obtain
$k=\psi=0$. Then, metric perturbations are described by $\eta$ and
$\chi$. By regularity they are $O(1)$ and $O(r^2)$ at the center,
respectively. The condition $k=0$ fixes the projection of
$\tilde{\xi}$ on $v^A$ completely, but $\psi=0$ fixes the orthogonal
part only up to a residual gauge freedom $\tilde{\xi}_A=f u_A$ where
the scalar $f$ obeys the equation $f'=\nu f$. This latter equation can
be thought of as an ODE in $r$ at constant $t$. We can give the
boundary value for this ODE at each moment of time, so the residual
gauge is an arbitrary function of time. We use it to set $\eta=0$ at
the center.

Using (\ref{regularity_eta}), (\ref{regularity_chi}) we define
\begin{equation}
\eta=\bar\eta , \qquad \chi=\phi+\eta=r^2\bar\chi ,
\end{equation}
where $\bar\chi$ is $O(1)$ at the center, but $\bar\eta$ is $O(x^2)$, due to
our gauge choice. The scalar field
perturbation $\Phi$ is already $O(1)$ and even at the center, compare
(\ref{regularity_Phi}). Equations (\ref{inveqbasis1}-\ref{inveqbasis3}) and 
(\ref{inveqwave}) are then
\begin{eqnarray}
  \frac{r}{a} \bar\chi' 
+ ( 1 + 2a^{-2} ) \bar\chi 
+ 4 Y^2 \left( \frac{\bar\eta}{r^2} - \bar\chi \right)
- \frac{4}{r} ( Y \dot\Phi + X \Phi' ) 
&=& 
0 , 
\label{constraint1_l0} \\
  \frac{1}{ar} \bar\eta' 
+ 4 Y^2 \left( \frac{\bar\eta}{r^2} - \bar\chi \right)
- \frac{4}{r} ( Y \dot\Phi + X \Phi' ) 
&=& 
0 , 
\label{constraint2_l0} \\
  \frac{r}{a} \dot{\bar\chi} 
- \frac{4}{r} ( X \dot\Phi + Y \Phi' ) 
&=& 
0 ,
\label{evoleq_l0} \\
- \ddot\Phi + \Phi'' 
- \frac{6}{r} a X Y \dot\Phi 
+ \frac{a}{2 r} ( 1 + 3 a^{-2} + 2 X^2 - 6 Y^2 ) \Phi' 
&& \nonumber \\
+ \frac{Y}{r} \dot{\bar\eta} 
+ a X ( 1 - 4 Y^2 ) \bar\chi 
+ \frac{4}{r^2} a X Y^2 \bar\eta 
+ 2 \left( \frac{\bar\eta}{r^2} - \bar\chi \right) r \dot{Y} 
&=& 
0 .
\end{eqnarray}
The last equation is the wave equation for the matter perturbation.
We do not have an evolution equation for $\bar\eta$. Instead, we have to 
calculate it by integration of the constraint (\ref{constraint2_l0}).
Finally $\bar\chi$ can be calculated from the evolution equation 
(\ref{evoleq_l0}) or by integration of the constraint (\ref{constraint1_l0}).

Again we rescale the variables. We also reorganize the variables to
eliminate $\dot\eta$ from the equations. (This is the same trick as
using $Y$ instead of $\dot\varphi$ to simplify the background equations.)
\begin{eqnarray}
u_1&\equiv &e^{-\tau} \left( \dot \Phi - \frac{Y}{r} \eta \right) , 
u_2 \equiv  e^{-\tau} \Phi' ,
u_3 \equiv  \Phi , 
\nonumber \\
u_4&\equiv & \bar\eta ,
u_5 \equiv  e^{-\tau} r \bar\chi .
\end{eqnarray}
Variables $(u_1,u_2,u_3,u_5)$ verify the following evolution equations:
\begin{equation}
\frac{\partial u}{\partial\tau}+A_4\frac{\partial u}{\partial x}+s=0 ,
\end{equation}
where the matrix $A_4$ is
\begin{equation}
A_4\equiv {\rm diag}(A_3,\lambda_0) ,
\end{equation}
and $s$ is the vector
\begin{eqnarray}
s_1 &=& 
  u1 
- \alpha \left[ - \frac{6a}{xe^{\xi_0}}  X Y u_1 
                + \left(\frac{3}{2}\bar{v}
                       +\frac{a}{xe^{\xi_0}}(\frac{1}{2}+X^2-3Y^2)
                  \right) u_2 
                + ( - 2 a \frac{X Y^2}{x^2 e^{2\xi_0}} 
                    + e^{-\tau}\frac{\dot{Y}}{x e^{\xi_0}} ) u_4
         \right.
\nonumber \\
&&       \left.
\hspace{1.5cm}    
+ \left(\frac{aX}{xe^{\xi_0}}(1-4Y^2)-2e^{-\tau}\dot{Y}\right) 
                       u_5 \right] ,
\\
s_2 &=& 
  u_2
- \alpha \left[ \left(\frac{a x e^{\xi_0}}{2}\bar{V}_0 
                     +\frac{a}{x e^{\xi_0}}(X^2+5Y^2)
                \right) u_1
              + \frac{2 a X Y}{x e^{\xi_0}} u_2
              + \left( 2 a \frac{X^2 Y}{x^2 e^{2\xi_0}} 
                  + e^{-\tau} \frac{\dot{X}}{x e^{\xi_0}} \right) u_4
              + \frac{4 a Y^3}{x e^{\xi_0}} u_5
         \right] ,
\\
s_3 &=&
- \alpha \left[ u_1 + \frac{Y}{x e^{\xi_0}} u_4 \right] ,
\\
s_5 &=&
  u_5
- \alpha \left[ \frac{4 a}{x e^{\xi_0}} ( X u_1 + Y u_2 ) 
                + 4 a \frac{X Y}{x^2 e^{2\xi_0}} u_4 
         \right] .
\end{eqnarray}
There are three constraints:
\begin{eqnarray}
\frac{\partial u_3}{\partial x} & = & a e^{\xi_0} u_2
, \\
\frac{\partial u_4}{\partial x} & = & c_4
, \\
\frac{\partial u_5}{\partial x} + \frac{b_5}{x} u_5 & = & c_5
,
\end{eqnarray}
where
\begin{eqnarray}
c_4 & =& 4 a^2 e^{\xi_0} ( Y u_1 + X u_2 + Y^2 u_5 ) ,\\
b_5 & =& 1 + a^2 ( 1 - 4 Y^2 ) ,\\
c_5 & =& \frac{4 a^2}{x} ( Y u_1 + X u_2 ) .
\end{eqnarray}
Note that we have a constraint, but no evolution equation, for $u_4$.
We have in fact constraints for $u_3$ and both $u_4$ and $u_5$,
so that the only degrees of freedom are those of a wave equation.
We obtain $u_4$ by solving a constraint at each time step, starting
from the gauge condition $u_4=0$ at $x=0$.


\subsection{Special case $l=1$}


For $l=1$ a general even perturbation is described by the objects 
($k_{AB},k,T_{AB},T_A,T^3$), which are only partially gauge-invariant: 
under an arbitrary coordinate transformation
generated by the vector 
$\xi_\mu dx^\mu=\tilde{\xi}_A Y dx^A +r^2\xi Y_{:a}dx^a$ 
these objects change as:
\begin{eqnarray}
k_{AB} &\longrightarrow& k_{AB} + (r^2\xi_{|A})_{|B} + (r^2\xi_{|B})_{|A} , \\
k      &\longrightarrow& k + 2\xi + (r^2)^{|A} \xi_{|A} , \\
T_{AB} &\longrightarrow& T_{AB} + r^2 ( t_{AB|C} \xi^{|C} +
                         t_{AC} {\xi^{|C}}_B + t_{BC} {\xi^{|C}}_A ) , \\
T_A    &\longrightarrow& T_A + r^2 ( t_{AB} \xi^{|B} - Q\xi_{|A} ) , \\
T^3    &\longrightarrow& T^3 + 2 Q\xi + (r^2Q)^{|A}\xi_{|A} .
\end{eqnarray}
We see that there is invariance under the $\tilde{\xi}_A$ part of the
transformation.  Therefore we have to impose just one partial gauge
condition.  The most interesting gauge condition is $k=0$, because
then we can eliminate all second derivatives from equations
(\ref{inveqbasis1}-\ref{inveqbasis7}).  Now matter perturbations are
described by $\eta,\psi,\chi$, which are $O(r)$, $O(r^2)$ and $O(r^3)$
at the center, respectively.  The condition $k=0$ does not fix the
gauge completely, and again we have a residual gauge freedom of
functions $\xi$ obeying equation $r^2v\xi'+\xi=0$. We use this freedom to
set $\eta\sim O(r^3)$ at the center.

Using (\ref{regularity_eta}), (\ref{regularity_psi}), (\ref{regularity_chi})
and (\ref{regularity_Phi}) we define
\begin{equation}
\eta=r\bar\eta , \qquad \psi=r^2 \bar\psi , \qquad 
\chi=\phi+\eta=r^3\bar\chi , \qquad \Phi=r \bar\Phi .
\end{equation}
where the barred variables are even and $O(1)$ at the center, except
$\bar\eta$, which is $O(r^2)$, due to our gauge choice.  Equations
(\ref{inveqbasis1}-\ref{inveqbasis6}) and (\ref{inveqwave}) are then
\begin{eqnarray}
  \frac{r}{a} \bar\chi' 
+ ( 2 + 3a^{-2} - 4Y^2 ) \bar\chi 
+ 4 Y^2 \frac{\bar\eta}{r^2}
- \frac{4}{r} \left( \frac{X}{a r} \bar\Phi 
                   + Y \dot{\bar\Phi} 
                   + X \bar\Phi' 
              \right) 
&=& 
0 , 
\\
  \frac{1}{ar} \bar\eta' 
+ 2 ( 1 - a^{-2} + X^2 + Y^2 ) \frac{\bar\eta}{r^2} 
+ ( a^{-2} - 2 X^2 - 2 Y^2 ) \bar\chi
- \frac{4}{r} X Y \bar\psi 
+ \frac{8}{a r^2} X \bar\Phi 
&=& 
0 , 
\\
  \frac{r}{a} \bar\psi' 
+ ( 2 + a^{-2} + 2 X^2 - 2 Y^2 ) \bar\psi 
+ 4 r X Y \left( \bar\chi - \frac{\bar\eta}{r^2} \right)
+ 4 \left( \frac{3 Y}{a r} \bar\Phi + X \dot{\bar\Phi} + Y \bar\Phi' \right) 
&=& 
0 , 
\\
  \frac{r}{a} \dot{\bar\chi} 
- ( 1 - 4 Y^2 ) \frac{\bar\psi}{r}
- \frac{4}{r} \left( \frac{Y}{a r} \bar\Phi + X \dot{\bar\Phi} + Y \bar\Phi' 
              \right) 
&=& 
0 , 
\\
  \frac{r}{a} \dot{\bar\psi} 
- 4 X Y \bar\psi 
+ ( 3 ( 1 - a^{-2} ) + 2 ( X^2 - Y^2 ) ) \frac{\bar\eta}{r}
- (     1 - a^{-2}   + 2 ( X^2 - Y^2 ) ) r \bar\chi
+ 4 \left( \frac{3 X}{a r} \bar\Phi + Y \dot{\bar\Phi} + X \bar\Phi' \right)
&=& 0 , 
\\
- \ddot{\bar\Phi} + \bar\Phi'' 
- \frac{2}{r} a X Y \dot{\bar\Phi}
+ ( 1 + 7 a^{-2} + 2 X^2 + 2 Y^2 ) \frac{a}{2 r} \bar\Phi'
+ \left( - V_0 + 8 \frac{Y^2}{r^2} \right) \bar\Phi 
&& \nonumber \\
+ \frac{Y}{r} \dot{\bar\eta} 
+ ( a X - 2 r \dot{Y} ) \bar\chi 
+ ( a X + 2 r \dot{Y} ) \frac{\bar\eta}{r^2}
+ \left( \frac{a Y}{r} ( 1 - 3 a^{-2} - 2 X^2 + 2 Y^2 ) - 2 \dot{X} \right)
     \bar\psi
&=& 
0 .
\end{eqnarray}

Again we rescale and regroup the variables:
\begin{eqnarray}
u_1&\equiv &e^{-2\tau} \left( \dot {\bar\Phi} - \frac{Y}{r} \eta \right) , 
u_2 \equiv  e^{-2\tau} \bar\Phi' ,
u_3 \equiv  e^{-\tau} \bar\Phi , 
\nonumber \\
u_4&\equiv &e^{-\tau} \bar\eta ,
u_5 \equiv  e^{-2\tau} r \bar\chi ,
u_6 \equiv  - e^{-2\tau} \bar\psi .
\end{eqnarray}
The variables $(u_1,u_2,u_3,u_5,u_6)$ obey the following evolution equations:
\begin{equation}
\frac{\partial u}{\partial\tau}+A_5\frac{\partial u}{\partial x}+s=0 ,
\end{equation}
where the matrix $A_5$ is
\begin{equation}
A_5\equiv {\rm diag}(A_3,\lambda_0,\lambda_0) ,
\end{equation}
and $s$ is the vector
\begin{eqnarray}
s_1 &=& 
  2 u1 
- \alpha \left[ - \frac{2 a X Y}{x e^{\xi_0}} u_1 
                + \left(\frac{7}{2}\bar{v}
                       +\frac{a}{x e^{\xi_0}}(\frac{1}{2}+X^2+Y^2)
                  \right) u_2 
                + \left( -\bar{V}_0 + \frac{8 Y^2}{x^2 e^{2\xi_0}} \right) u_3
                + \left( a \frac{X}{x^2 e^{2\xi_0}} (1 - 2 Y^2) 
                    + e^{-\tau} \frac{\dot{Y}}{x e^{\xi_0}} \right) u_4
         \right.
\nonumber \\ &&
\hspace{1.5cm} 
         \left.
                + \left(\frac{a X}{x e^{\xi_0}}-2 e^{-\tau}\dot{Y}\right) u_5
                + \left( 3 \bar{v} Y - \frac{a Y}{x e^{\xi_0}} (1-2X^2+2Y^2)
                       +2 e^{-\tau} \dot{X} \right) u_6
         \right] ,
\\
s_2 &=& 
  2 u_2
- \alpha \left[ \left(\frac{a x e^{\xi_0}}{2}\bar{V}_0 
                     +\frac{a}{x e^{\xi_0}}(X^2+Y^2)
                \right) u_1
              - \frac{2 a X Y}{x e^{\xi_0}} u_2
              - \frac{8 X Y}{x^2 e^{2\xi_0}} u_3
              + \left( - 2 a Y ( \bar{V}_0 + \frac{Y^2}{x^2 e^{2\xi_0}} )
                 + e^{-\tau}\frac{\dot X}{x e^{\xi_0}} \right) u_4
         \right.
\nonumber \\ &&
\hspace{1.5cm} 
         \left.
              + \frac{Y}{x e^{\xi_0}}\left(-\frac{1}{a}+2a(X^2+Y^2)\right) u_5
              - \frac{4 a X Y^2}{x e^{\xi_0}} u_6
         \right] ,
\\
s_3 &=&
  u_3
- \alpha \left[ u_1 + \frac{Y}{x e^{\xi_0}} u_4 \right] ,
\\
s_5 &=&
  2 u_5
- \alpha \left[ \frac{4 a}{x e^{\xi_0}} ( X u_1 + Y u_2 )
              + \frac{4}{x^2 e^{2\xi_0}} Y u_3
              + 4 a \frac{X Y}{x^2 e^{2\xi_0}} u_4 
              - \frac{a}{x e^{\xi_0}} (1-4Y^2) u_6
         \right] ,
\\
s_6 &=&
  2 u_6
- \alpha \left[ \frac{4 a}{x e^{\xi_0}} ( Y u_1 + X u_2 )
              + \frac{12}{x^2 e^{2\xi_0}} X u_3
              + ( 3a \bar{V}_0 + \frac{2a}{x^2 e^{2\xi_0}} (X^2 +  Y^2) ) u_4
         \right.
\nonumber \\ &&
\hspace{1.5cm}
         \left.
              - \left(  a x e^{\xi_0} \bar{V}_0 
                  + \frac{2 a}{x e^{\xi_0}}(X^2-Y^2) \right) u_5
              + \frac{4 a}{x e^{\xi_0}} X Y u_6
         \right] .
\end{eqnarray}
There are four constraints:
\begin{eqnarray}
\frac{\partial u_3}{\partial x} & = & a e^{\xi_0} u_2
, \\
\frac{\partial u_4}{\partial x} + \frac{b_4}{x} u_4 & = & c_4
, \\
\frac{\partial u_5}{\partial x} + \frac{b_5}{x} u_5 & = & c_5
, \\
\frac{\partial u_6}{\partial x} + \frac{b_6}{x} u_6 & = & c_6
,
\end{eqnarray}
where
\begin{eqnarray}
b_4 & =& -2 + 2 a^2 ( 1 + X^2 + Y^2 ) , \\
b_5 & =& 2 + 2 a^2 ( 1 - 2 Y^2 ) , \\
b_6 & =& 1 + 2 a^2 ( 1 + X^2 -Y^2 ) , \\
c_4 & =& - \frac{8 a X}{x} u_3 
         + e^{\xi_0} ( -1 + 2 a^2 ( X^2 + Y^2 ) ) u_5
         - 4 e^{\xi_0} a^2 X Y u_6 , \\
c_5 & =& \frac{4 a^2}{x} ( Y u_1 + X u_2 + \bar{v} X u_3 ) , \\
c_6 & =& \frac{4 a^2}{x} ( X u_1 + Y u_2 + 3 \bar{v} Y u_3 + X Y u_5 ) .
\end{eqnarray}
Note that again we do not have an evolution equation for $u_4$, and
that we have constraints for all variables other than $u_1$ and $u_2$,
so that the only degrees of freedom are those of a wave equation.
There is a consistency condition at the center:
\begin{equation}
u_6 - { 4\over e^{\xi_0}} {dY\over dx} u_3=O(x^2) .
\end{equation}
Again we impose $u_4=0$ at $x=0$ as a gauge condition.


\section{Numerical results}


Our numerical code, and the tests we have performed, are described in
an appendix. Here we only summarize three important points. 

The code treats the boundaries $x=0$ (center of spherical symmetry)
and $x=1$ (boundary of domain of dependence) in exactly the same way
as all other points. On a flat empty background spacetime, it is
second-order convergent, the origin $x=0$ is stable, and waves cleanly
leave the computational domain at $x=1$ without numerical backscatter.

On the Choptuik background we observe second-order convergence for
most values of $x$ and $\tau$. Convergence of a lower than second
order is observed near $x=0$, twice per period in $\tau$. These are
the values of $\tau$ where certain coefficients of the background
solution change rapidly in time, namely at the minima and maxima of
the background scalar field. A typical solution (as we shall discuss
below) is an exponentially damped quasiperiodic oscillation.
Convergence inevitably breaks down at large $\tau$ for two reasons:
the oscillations at different numerical resolutions gradually drift
out of phase, and small differences in the exponential decay rates at
different resolutions have a cumulative effect on the amplitude.

As we discuss in detail in the appendix, the numerical code has a
subtle instability which becomes apparent only at high $l$ at high
resolutions. The instability is already present in the free wave
equation (in self-similar coordinates) on Minkowski space. We have
found a way of repairing it in Minkowski space, but it persists on the
Choptuik background. At low resolution this instability can be
neglected, and we see convergence up to a resolution of $\Delta
x=1/800$. 

In spite of the inevitable absence of pointwise convergence at late
times, and in spite of the numerical instability, our main result
appears secure: all non-spherical physical perturbation modes, for all
initial data, decay exponentially in $\tau$.  The exponential decay is
typically rapid. Only for even $l=2$ perturbations is the decay quite
slow, but there (as for low $l$ in general) we have good convergence
of the solution itself, and therefore the decay exponent.

Due to the discrete self-similarity of the background solution, the
perturbations decay in a complicated fashion, with the exponential
decay apparent only over many periods. The
background-dependent coefficients of the perturbation equations are
periodic in $\tau$ (at constant $x$).  Therefore the general form of
the perturbation is a sum of terms of the form 
\begin{equation}
u(x,\tau)={\rm Re}\left[C\,e^{\lambda\tau} f(x,\tau)\right],
\end{equation}
with $C$, $\lambda$ and $f(x,\tau)$ all complex, and
$f(x,\tau+\Delta)=f(x,\tau)$. Once the most slowly decaying mode
dominates, only one such term is left. In real notation, it is
\begin{equation}
\label{quasiperiodic}
u(x,\tau) = e^{\kappa\tau}\left[C_1 \cos(\omega\tau) f_1(x,\tau) + C_2
\sin(\omega\tau) f_2(x,\tau)\right], 
\end{equation}
with $\kappa={\rm Re}\lambda$, $\omega={\rm Im}\lambda$, $C_1$, $C_2$,
$f_1$ and $f_2$ now real, and $f_1$ and $f_2$ again periodic.  This
means that $u(x,\tau)$, even after the exponential decay has been
taken out, is not periodic in $\tau$ unless $\omega$ is commensurate
with $2\pi/\Delta$. Furthermore, $C_1$ and $C_2$, and in particular their
ratio, depend on the perturbation initial data. Therefore, the
complex exponent $\lambda$ is not easy to read off. Nevertheless, to the
extent to which they are approximated by (\ref{quasiperiodic}), the
Fourier transform in $\tau$ of the data with the exponential decay
taken out should be peaked around the set of frequencies
\begin{equation}
\label{quasiperiodic_spectrum}
N{2\pi\over\Delta} + \omega
\end{equation}
for integer $N$. The background is not only periodic in $\tau$ with
period $\Delta$, but has an additional symmetry. The background scalar
field obeys $\varphi(\tau+\Delta/2,x)=-\varphi(\tau,x)$, while the
background metric coefficients obey
$g(\tau+\Delta/2,x)=g(\tau,x)$. The perturbations inherit this
additional symmetry. Therefore, in the spectrum
(\ref{quasiperiodic_spectrum}) of the scalar field perturbations $u_1$
to $u_3$, only odd integers $N$ appear, while in the spectrum of the
metric perturbations $u_4$ to $u_8$, only even value of $N$
appear. This must be taken into account when we read off $\omega$ from
the spectrum. Because the $N$ are either even or odd,
$\omega\Delta/2\pi$ is defined modulo $2$ (and not modulo $1$ as one
might expect), and we define it to be $0\le\omega\Delta/2\pi<2$. For
example, with the highest peak in the spectrum of $u_1$ at
$\omega\Delta/2\pi=5.3$, and the highest peak in the spectrum of $u_4$
at $\omega\Delta/2\pi=6.3$, we consistently obtain
$\omega\Delta/2\pi=0.3$.

The only exception to this complicated behavior are the spherical
($l=0$) perturbations. At large $\tau$ they are dominated by a single
growing mode with $\lambda$ real. (The fact that there is a single
growing mode is of course at the center of critical phenomena in
critical collapse, and this unique $\lambda$ must then be real because
the background is real.) Here we can read off both $\kappa$ and
$f(x,\tau)$ quite clearly. We find $\kappa\Delta=9.21$. This
corresponds to $\kappa=2.67$, and a critical exponent for the black
hole mass of $\gamma=1/\kappa=0.374$. This agrees to all three digits
with the value of the critical exponents obtained from collapse
simulations \cite{Choptuik}, and a perturbative calculation
\cite{Gundlach} that is completely independent from the present one.

For $l>0$, we have obtained estimates of $\kappa$ and $\omega$ by
first adjusting the value of $\kappa$ until the rescaled $u_{\rm
rescaled}\equiv e^{-\kappa\tau} u$ appeared to be neither increasing
nor decreasing over many periods. The resulting $u_{\rm rescaled}$ is
then quasiperiodic. We have carried out a discrete Fourier transform
on the time series $u_{1,\rm rescaled}(0,\tau)$ over a range of
$10\Delta$. The result has sharp peaks spaced at intervals
$4\pi/\Delta$ due to the additional symmetry in the background mentioned
above.  In the
special cases $l=0$ and $l=1$ the function $u_{1,\rm rescaled}$ is
clearly periodic ($\omega=0$), and the line spectrum is very
sharp. 

\begin{figure}
\epsfysize=12cm \epsfbox{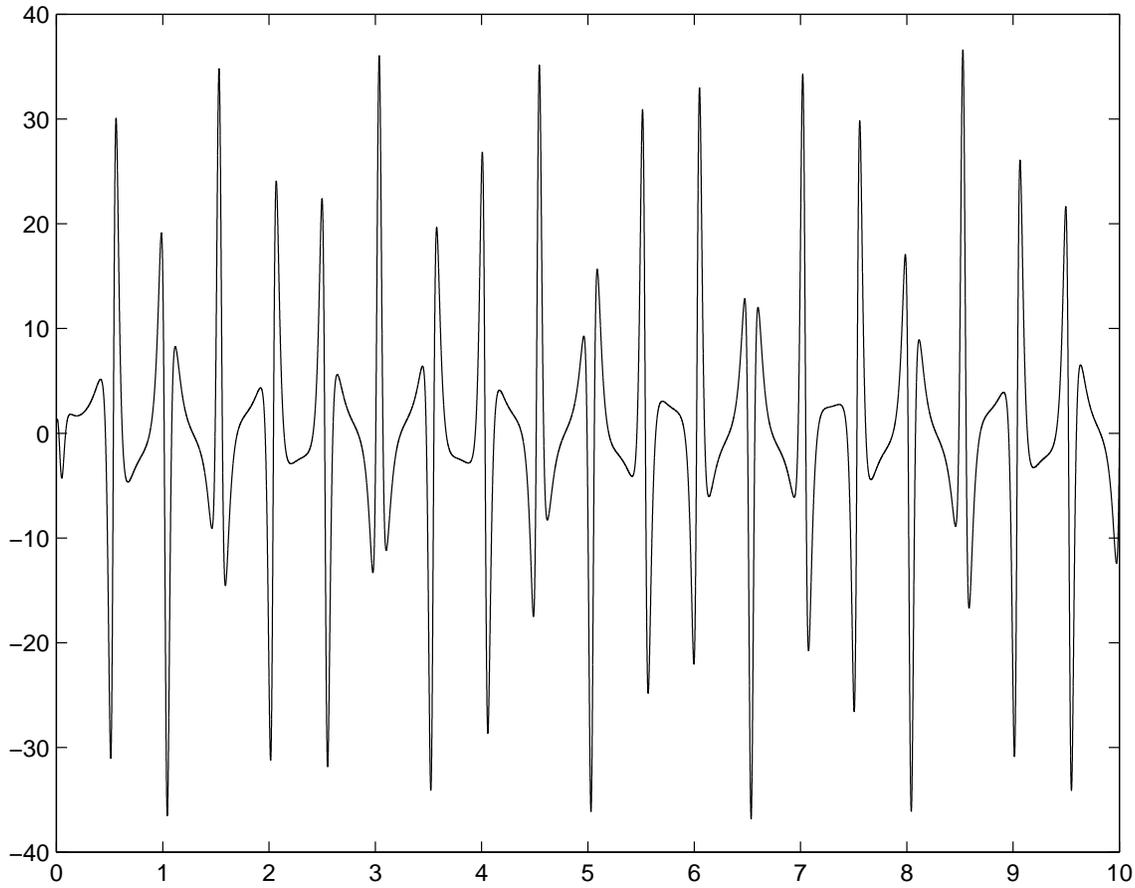}
\caption{$u_1$ versus $\tau$ at $x=0$. An overall exponential decay
has been compensated for. The scale on the vertical axis
is irrelevant, as the equations are linear. On the horizontal axis we
have marked background periods, that is $\tau/\Delta$. The sharp peaks
are typical features. Although it is not clear from this plot, they
are perfectly smooth.} 
\label{fig:even_l=2_rescaled}
\end{figure}

\begin{figure}
\epsfysize=12cm \epsfbox{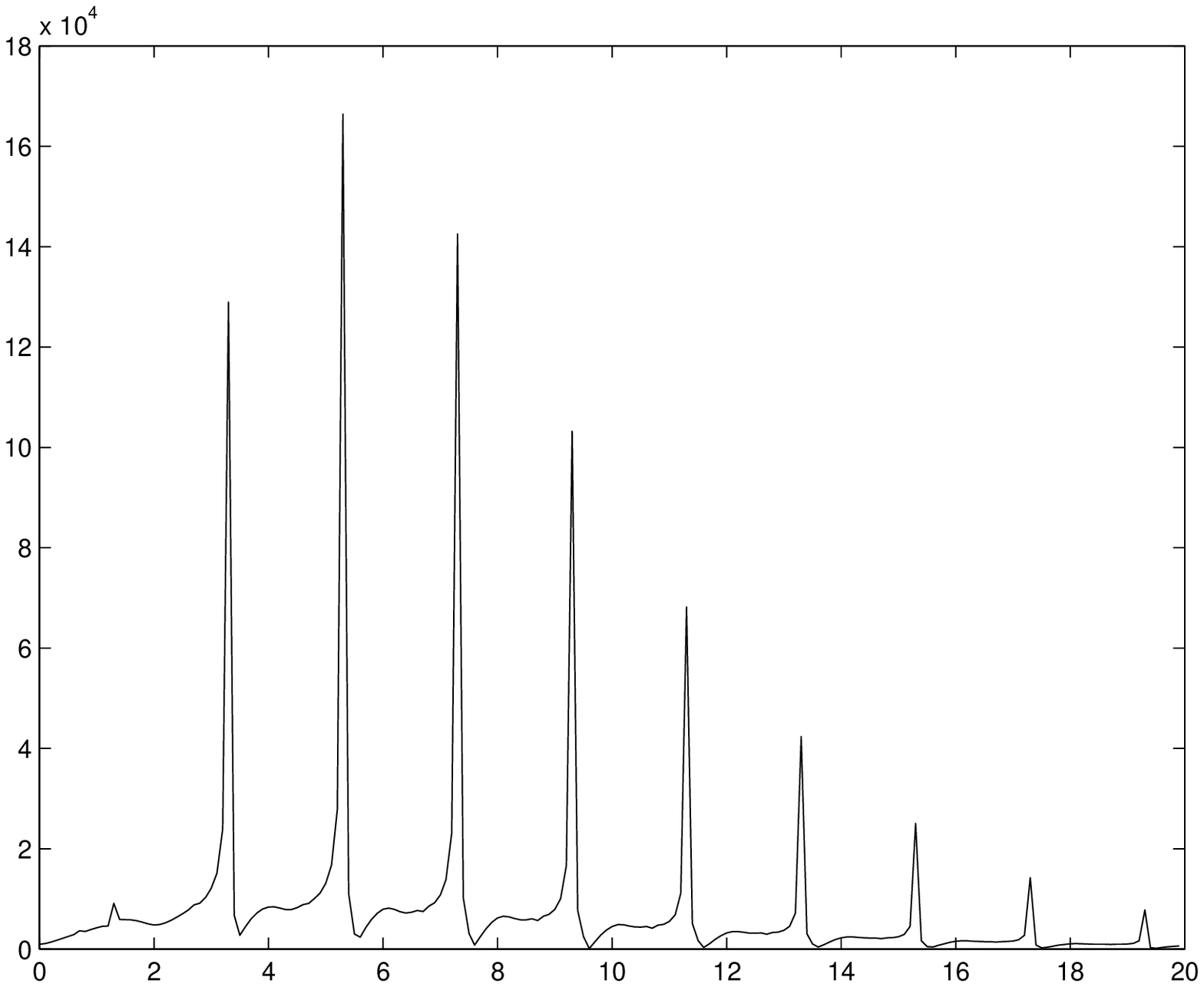}
\caption{The low-frequency end of the discrete Fourier transform of
the previous figure. The vertical scale is again irrelevant. On the
horizontal axis we have marked frequency in units of the background
frequency, that is $(\omega\Delta)/(2\pi)$. The quasiperiodic nature
of the signal shows up in the peaks situated at
$(\omega\Delta)/(2\pi)=1.3, 3.3, 5.3, \dots$. The spectrum decays
rapidly at high frequencies.}
\label{fig:even_l=2_fft}
\end{figure}

The estimated values of $\kappa$ and $\omega$ are tabulated for
different resolutions in Table 1. As an example, we show the value of
$u_1$ for even $l=2$ perturbations at $x=0$ as a function of $\tau$,
after an exponential decay has been taken out, in
Fig.~\ref{fig:even_l=2_rescaled}. In Fig.~\ref{fig:even_l=2_fft} we
show the low frequency part of the discrete Fourier transform of
Fig.~\ref{fig:even_l=2_rescaled}. The quasiperiodic nature of the
signal becomes clear in that there is a series of peaks obeying
(\ref{quasiperiodic_spectrum}).

As the background spacetime is periodic in $\tau$, and the
perturbation equations are linear, evolving the perturbations for one
period is equivalent to multiplying them by a transfer matrix. For odd
perturbations, this matrix has size $(2N)^2$, and for even
perturbations $(4N)^2$, where $N$ is the number of grid points in $x$,
and two and four respectively is the number of degrees of freedom. For
$N=50$ and $100$, we have verified that the logarithm of the largest
eigenvalue of the transfer matrix agrees with $\lambda\Delta$. These
matrices contain of course all the information that there is about the
system, but for larger $N$ the computation time and memory requirement
for calculating these matrices and their eigenvalues quickly becomes
prohibitive, scaling as $N^4$. However, if we use generic initial
data, in which no $u$ vanishes at any $x$ (except odd $u$ at $x=0$),
we have a mixture of all perturbation modes, and at late enough times
the most slowly decaying mode has taken over. 


With increasing $l$, the even parity numerical code appears to be more
and more sensitive to small errors in the background solution, as the
solutions obtained at different resolutions drift apart more and more
rapidly. The solution at late times depends sensitively on the initial
data, so that the system looks chaotic.  This problem may be
unavoidable with any code. Our results still seem to capture the
correct overall behavior, as the values of $\kappa$ and $\omega$
obtained at different resolutions differ much less than the actual
time series. We believe that the explanation is that different
resolutions agree reasonably well on the periodic functions $f_1$ and
$f_2$, but that the initial data-dependent coefficients $C_1$ and
$C_2$ take essentially random values at late times for different
resolutions.

In summary, we find that both even and odd perturbations decay
exponentially for all physical values of $l$. It is clear that
perturbations with large $l$ will decay more and more quickly because
of the presence of the terms $u_{,\tau}=-lu+\dots$ in all evolution
equations. (These terms are introduced by the scaling of perturbations
with $r^l$ to keep them regular at $r=0$.) The numerical evolutions
confirm that higher $l$ modes decay more and more rapidly. We can
therefore affirm that all values of $l$ decay, even though we have
checked this explicitly only for the lowest few values. The most
slowly decaying mode occurs in the $l=2$ polar perturbations, with
$\lambda\simeq-0.07\times(1/\Delta)+0.3\times (2\pi i/\Delta) \simeq
-0.02 + 0.55i$. As this mode decays so very slowly, there may be an
intermediate range of $p-p_*$ for a given one-parameter family of
initial data where this perturbation becomes universally visible.  For
$p-p_*$ small enough, however, the spherical universal solution will
again dominate.


\acknowledgments

We would like to thank M. Alcubierre, M. Choptuik, A. Dom\'\i nguez,
D. Garfinkle, V. Moncrief, A. Rendall and P. Walker for helpful
conversations.  JMM would like to thank the Albert-Einstein-Institut
for hospitality.  JMM was supported by the 1994 Plan de Formaci\'on de
Personal Investigador of the Comunidad Aut\'onoma de Madrid.


\begin{table}
\begin{tabular}{lddddd}
System & & & $\left(\kappa\Delta,{\omega\Delta/2\pi}\right)$ & & \\
grid points & 100 & 200 & 400 & 800 & 1600 \\
\hline
even $l=0$ & 9.24, 0.0  & 9.21, 0.0  & 9.21, 0.0  & 9.21, 0.0 & 9.21, 0.0\\
even $l=1$ & -0.34, 0.0 & -0.31, 0.0 & -0.48, 0.0 & noisy &
-0.30, 0.0\\
\hline
even $l=2$ & -0.08, 0.3 & -0.07, 0.3 & -0.06, 0.3 & -0.07, 0.3 & -0.07 0.3 \\
even $l=3$ & -1.63, 1.6 & -1.65, 1.6 & -1.65, 1.6 & -1.65, 1.6 & -1.66, 1.6 \\
even $l=4$ & -2.8, 0.9  & -2.9, 0.9  & -2.9, 0.9  & -3.0, 0.9  &
noisy \\
even $l=5$ & -4.0, 0.2 & -4.25, 0.2 & -3.9, 0.2 & -3.65, 0.3 & noisy \\
\hline
odd $l=2$ & -2.20, 1.9 & -2.28, 1.9 & -2.30, 1.9 & -2.30, 1.9 & -1.8, 3.0\\
odd $l=3$ & -3.13, 1.3 & -3.23, 1.4 & -3.27, 1.4 & -3.28, 1.4 & noisy \\
odd $l=4$ & -4.05, 0.7 & -4.20, 0.7 & -4.25, 0.7 & -4.27, 0.7 &
noisy \\
odd $l=5$ & -5.0, 0.0 & -5.2, 0.0 & -5.2, 0.1 & -5.3, 0.1 & -5.3,
0.1 \\
\end{tabular}
\caption{Summary of eigenvalues $\lambda$, read off from
$u_1(x=0,\tau)$. The values of kappa were obtained by eliminating an
exponential factor. The values of $\omega$ were obtained from
a discrete Fourier transform of the result. Values of
$\omega\Delta/2\pi$ are defined modulo $2$, while peaks in the
discrete Fourier transform of $u_1$ are located at
$\omega\Delta/2\pi+1+2N$ for non-negative integer $N$. As we have
integrated over a range of $10\Delta$ in $\tau$, $\omega\Delta/2\pi$ can
only be estimated in multiples of $0.1$.  Results marked ``noisy''
are dominated by numerical error. In the Fourier transform this shows
up as high frequency noise.}
\label{table:runs}
\end{table}


\begin{table}
\begin{tabular}{lddddd}
System & & & $\left(\kappa\Delta,{\omega\Delta/2\pi}\right)$ & & \\
grid points & 100 & 200 & 400 & 800 & 1600 \\
\hline
even $l=2$ & 1.1, 0.3 & 0.15, 0.3 & 0.0, 0.3 & -0.05, 0.3 & -0.07, 0.3 \\
even $l=3$ & 0.25, 1.4 & -1.3, 1.6 & -1.55, 1.6 & -1.63, 1.6 & -1.65,
1.6 \\
even $l=4$ & -0.53, 0.5 & -2.0, 0.8 & -2.75, 0.9 & -2.8, 1.0 &
noisy \\
even $l=5$ & -1.45, 1.5 & -2.77, 0.0 & -3.3, 0.1 & -3.2, 0.2 &
noisy \\
\hline
odd $l=2$ & -1.74, 1.8 & -2.18, 1.9 & -2.30, 1.9 & -2.30, 1.9 &
noisy \\
odd $l=3$ & -1.95, 1.2 & -2.95, 1.3 & -3.2, 1.4 & -3.25, 1.4 &
noisy \\
odd $l=4$ & -2.2, 0.4 & -3.55, 0.7 & -4.1, 0.7 & -4.25, 0.7 & noisy
\\
odd $l=5$ & -2.7, 1.4 & -4.1, 1.9 & -5.0, 0.0 & -5.2, 0.1 & noisy \\
\end{tabular}
\caption{The same with an alternative finite differencing method for
$l\ge2$, defined by Eq. (\ref{newmethod}). Note that convergence is much
slower, but that for 800 grid points these results agree quite well
with those of the other method.}
\label{table:newruns}
\end{table}


\appendix


\section{Background solution}
\label{appendixA}


Following Choptuik, we introduce the scale-invariant and dimensionless
auxiliary fields for the spherically symmetric scalar field system as
\begin{equation}
X=\sqrt{2\pi }r\varphi^\prime , \qquad Y=\sqrt{2\pi }r\dot\varphi ,
\end{equation}
in the radial basis. (Do not confuse Y with a spherical harmonic.)
We use the dependent variables $g=a/\alpha$, $a$, $X$ and $Y$, and the
coordinates $x$ and $\tau$.
With the shorthand 
\begin{equation}
D\equiv  xge^{\xi_0}(1-\xi_{0,\tau}) ,
\end{equation}
the wave equation for $\varphi$ is equivalent to the first-order system
\begin{equation}
x\left(\matrix{X_{,x} \cr Y_{,x}}\right)=
\frac{1}{1-D^2}
\left(\matrix{1 & D \cr D & 1}\right)
\left(\matrix{-[\frac{1}{2}(1+a^2)+a^2(X^2-Y^2)]X+gxe^{\xi_0}Y_{,\tau} \cr
        \ \ \; [\frac{1}{2}(3-a^2)+a^2(X^2-Y^2)]Y+gxe^{\xi_0}X_{,\tau}}\right)
.
\end{equation}
The Einstein equations are
\begin{eqnarray}
xg_{,x}&=&g(1-a^2) , \\
xa_{,x}&=&\frac{a}{2}[1-a^2+2a^2(X^2+Y^2)] , \\
a_{,\tau}&=&-(1-\xi_{0,\tau})xa_{,x}+\frac{2a^3}{gxe^{\xi_0}} XY .
\label{a_constraint}
\end{eqnarray}
In order to exclude a conical singularity at $r=0$, we impose $a=1$ at
$x=0$. In order to fix the remaining coordinate ambiguity $t\to f(t)$
we impose $g=1$ at $x=0$. We make $x=1$ an ingoing null surface by
imposing $D=1$ at $x=1$. $\xi_0$ is not initially known, but is
determined together with the dynamical fields $X$, $Y$, $a$ and $g$ of
the critical solution.

We have recalculated the background using the numerical code of
Gundlach\cite{Gundlach}, slightly modified to use $x$ instead of 
$\zeta=\ln x$, which
results in a better treatment of $x=0$.  If the solution is regular,
$X$ and $Y$ vanish at $x=0$. Therefore we work with $X_2\equiv
x^{-2}X$ and $Y_1\equiv x^{-1}Y$.  $x=0$ and $x=1$ are regular
singular points of the equations. The regularity condition (vanishing
of the numerator in the wave equation) is
\begin{equation}
X_2 = {1\over 3}e^{\xi_0}\left[Y_{1,\tau}+(1-d\xi_0/d\tau)Y_1\right]
\end{equation}
at $x=0$, while at $x=1$ it is
\begin{equation}
  \left[ 1+a^2(1+2X_2^2-2Y_1^2)\right] X_2
+ \left[-3+a^2(1-2X_2^2+2Y_1^2)\right] Y_1
-2\left(1-{d\xi_0\over dx}\right)^{-1}
\left({\partial Y_1\over \partial \tau}
+ {\partial X_2\over \partial \tau} \right) = 0.
\end{equation}

The discrete self-similarity of the background is equivalent to
periodicity of $X$, $Y$, $a$ and $g$ in $\tau$, with a period $\Delta$
that is initially unknown. $a$ is treated as a functional of $X$, $Y$,
$g$ and $\xi_0$, by solving equation (\ref{a_constraint}), with
periodic boundary conditions in $\tau$ for each value of $x$. Note
that this equation is linear in $a^{-2}$. Periodicity is imposed by
representing $X$, $Y$, $g$ and $\xi_0$ through a (truncated) Fourier
series. $\tau$-derivatives are calculated, and equation
(\ref{a_constraint}) is solved, in Fourier space. This makes the
numerical method a pseudo-spectral one. The $y$-derivatives are
implemented through finite differencing on a grid equally spaced in
$x$, and are solved by relaxation, together with the algebraic and ODE
(pseudo-algebraic) boundary conditions at $x=0$ and $x=1$.

We have calculated the background solution using points 51, 101, ...,
1601 on the range $0\le x\le1$, always with 128 points per period
$0\le \tau <\Delta$. It was shown in
\cite{Gundlach} that this resolution in $\tau$ is large enough so that
numerical error is dominated by resolution in $x$ and systematic error
effects at $x=0$ and $x=1$.

\begin{figure}
\epsfysize=14cm \epsfbox{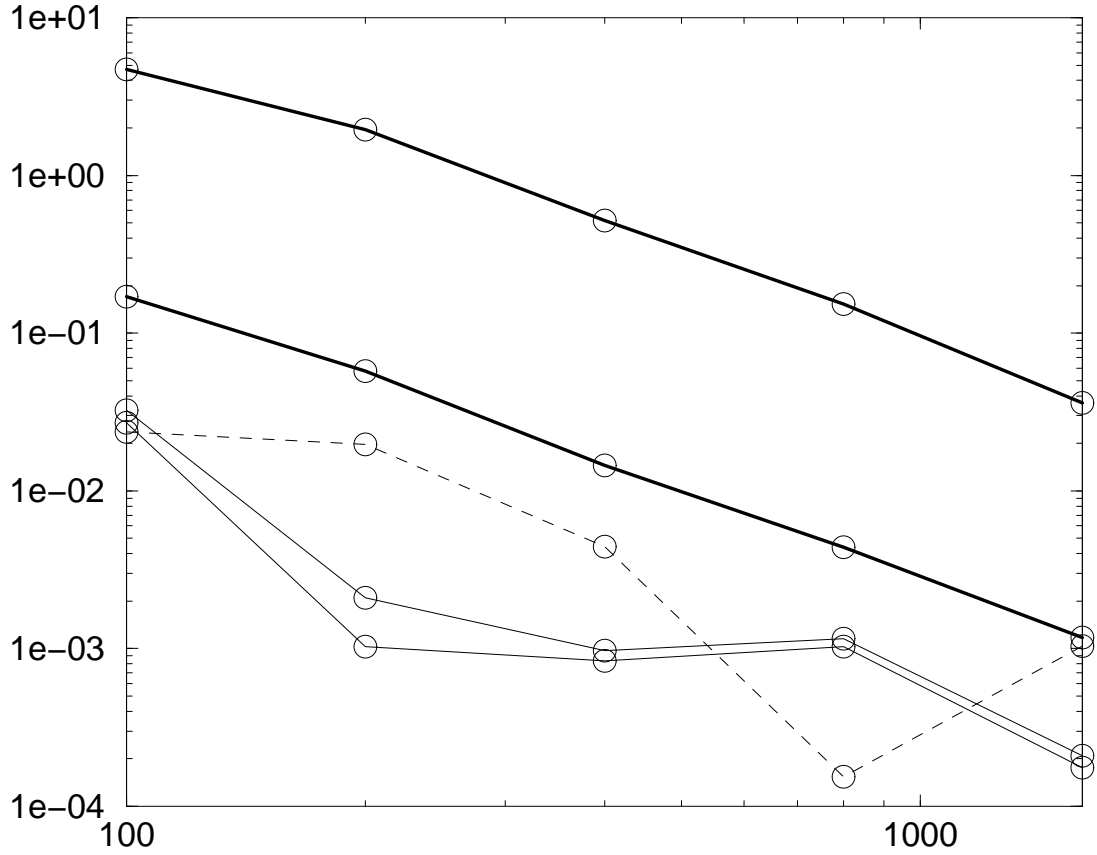}
\caption{Convergence of the background solution with increasing
resolution in $x$. On the horizontal axis we plot the logarithm of the
number of grid points, on the vertical axis the logarithm of the
difference between one numerical solution and the one with half the
resolution.  The two thick lines are the maximal and root-mean-squared
differences (over both $0\le x\le 1$ and one period in $\tau$) of all
fields. The thin lines are the differences of the ``eigenvalue''
$\xi_0(\tau)$. The dashed line is the difference in the eigenvalue
$\Delta$.}
\label{fig:bgconv}
\end{figure}

We observe second-order convergence, measured by the maximal and
root-mean-squared differences of $X_2$, $Y_1$, $a$ and $g$, from 51 to
1601 grid points in $x$ (with 128 Fourier components in $\tau$).
$\Delta$ and $\xi_0$ (in the maximum and root-mean-squared norms) also
show convergence, but not to a distinct order. This is illustrated in
Fig.~\ref{fig:bgconv}. For the perturbation calculations we have
always used the high-resolution background, downsampled as necessary.


\section{Perturbations numerical method}


The perturbation equations are of the form
\begin{equation}
{\partial u\over \partial \tau} + A {\partial u\over \partial x} + Bu =
0, 
\end{equation}
where $u$ is a vector of unknowns, and $A$ and $B$ are
background-dependent matrices.

As by definition $x=1$ is an ingoing spherical null surface, the
domain of dependence of perturbation data at $\tau=0$, $0\le x\le1$
for $\tau\ge0$ is precisely $0\le x\le 1$. Both scalar and
gravitational waves travel only from smaller to larger $x$ for
$x\ge1$. In order to implement a time evolution without an artificial
boundary condition at $x=1$, we use an evolution scheme that
explicitly uses the characteristic speeds and therefore changes over
smoothly to upwind $x$-derivatives for $x\ge 1$. The numerical
method we have used is similar to that used for the perturbations of
the perfect fluid critical solutions in a previous publication
\cite{critfluid,Leveque}, but is second order in space and time. It uses
second order one-sided derivatives in $x$, and is Runge-Kutta-like in
$\tau$:
\begin{eqnarray} 
\label{forward}
u^{n+{1\over2}}_j 
& = & 
  u^n_j 
- {\Delta \tau \over2} 
     \left[ (A_-)^n_j {4u^n_{j+1} - u^n_{j+2}- 3 u^n_j \over2 \Delta x} 
          - (A_+)^n_j {4u^n_{j-1} - u^n_{j-2}- 3 u^n_j \over2 \Delta x} 
          + B^n_j u^n_j 
     \right]
, \\ 
\label{leapfrog}
u^{n+1}_j 
& = & 
  u^n_j 
- \Delta \tau 
    \left[ (A_-)^{n+{1\over2}}_j 
              {4u^{n+{1\over2}}_{j+1}-u^{n+{1\over2}}_{j+2}-3 u^{n+{1\over2}}_j 
               \over2 \Delta x} 
         - (A_+)^{n+{1\over2}}_j 
              {4u^{n+{1\over2}}_{j-1}-u^{n+{1\over2}}_{j-2}-3 u^{n+{1\over2}}_j 
               \over2 \Delta x} 
         + B^{n+{1\over2}}_j u^{n+{1\over2}}_j 
    \right]
.
\end{eqnarray}
Here $A_++A_-=A$.  In order to use the characteristic speeds in the
finite differencing scheme, it is necessary to split $A$ into a sum
over its eigenvalues according to their sign, so that, for example,
\begin{equation}
A_{3+}= \lambda_+ 
\left(\matrix{ {1/2} & -{1/2} & \cr
-{1/2} & {1/2} & \cr
& & 0 \cr}\right)
+ \lambda_0
\left(\matrix{0 & & \cr & 0 & \cr & & 1 \cr}\right) ,
\qquad
A_{3-}=\lambda_-
\left(\matrix{ {1/2} & {1/2} & \cr
{1/2} & {1/2} & \cr
& & 0 \cr}\right) 
\end{equation}
For $x\ge1$, $\lambda_-$ becomes positive, so that$A_+=A$ and $A_-=0$,
so that we do not need the downwind derivative there. We go just one
grid point beyond $x=1$, so that the last grid point just before $x=1$
still has two points to its right in order to take a right derivative
there. All grid points further to the right only require left
derivatives. This means that we could have extended the numerical
domain to large $x$. We have chosen $0\le x \le 1+\Delta x$ because it is
the smallest numerical domain in which we stay in the domain of
dependence of the perturbation initial data for all $\tau$, 
while using a one-sided
three-point stencil. (If we used a first-order differencing scheme,
with two-point stencils, the numerical domain $0\le x\le 1$
would be sufficient.)

We might refer to the method just described as the second-order
characteristic method. It is explicit and second order. One obtains an
implicit method by averaging $u^n$ and $u^{n+1}$ to obtain a new
improved estimate for $u^{n+1/2}$, 
\begin{equation}
\label{average}
u^{n+1/2} = (u^n + u^{n+1}) / 2,
\end{equation}
and iterating the pair (\ref{average},\ref{leapfrog}) of equations
until $u^{n+1}$ has converged. Let us call this the iterated
characteristic method. 

The boundary $x=0$ does not require special treatment, as $u(-x)=\pm
u(x)$ for all $u$, so that ghost grid points with $x<0$ are available
for taking derivatives. The one-sided differencing operators do not
give exactly zero at $x=0$ even if analytically $\partial u/\partial
x(0)=0$, but that is consistent: all terms in the finite difference
equations combine so that $u(0)$ remains zero to machine
precision if $u$ is odd initially. This also ensures that source terms
of the form $u/x$ for odd $u$ in the evolution equations are well
behaved numerically.

Another promising candidate for numerical treatment
of the equations is the iterated Crank-Nicholson method. In this
method, we need to treat the boundaries specially. At $x=0$ we have
tried updating the boundary point by extrapolation from its 
next neighbors  at each iteration, taking into account
that $u$ is either odd in $x$ (and vanishes at $x=0$) or even in
$x$. Alternatively we have used the exact value $\partial u/\partial
x=0$ for even $u$ and the finite difference stencil $\partial
u/\partial x=[u(\Delta x)-u(0)]/\Delta x$, which is second order at $x=0$ for
odd $u$. At $x=1+\Delta x$, we have either used extrapolation, or the
one-sided (left side only) second-order stencil of the characteristic
methods. 

The constraints are solved by integrating from
$x=0$ out to $x=1$. For stability, we do not evolve any variable for
which we have a constraint, but instead calculate it from the
constraints, including at the half-step $n+{1\over2}$. For simple
integrations $du/dx=c$ with $u$ an even function of $x$, and $u(0)$
given, we use the trapezoidal rule
\begin{equation}
u_{j+1} = u_j + {\Delta x\over 2}\left(c_j + c_{j+1}\right). 
\end{equation}
For the ODE $du/dx+bu/x=c$, with $b$ and $c$ even in $x$ and,
therefore, $u$ odd in $x$, we use
\begin{equation}
u_{j+1} = u_j + \left(1 + {\Delta x\, b_{j+1} \over x_j+x_{j+1}}\right)^{-1}
\left[\left(1 - {\Delta x\, b_j \over x_j+x_{j+1}}\right) + {\Delta x\over 2}
(c_j+c_{j+1})\right].
\end{equation}
This scheme is second-order accurate at all grid points. For those
variables $u$ that are even in $x$ and of $O(1)$ at $x=0$, because
$c$ is odd and of $O(x^{-1})$, we use the same scheme, with
coefficients $b-1$ and $cx$, in order to first calculate the odd
function $\bar u=ux$. Then we divide by $x$ to obtain $u$. In this case we
extrapolate twice: first $cx$ to $x=0$, and then $\bar u/x=u$ to
$x=0$. 

The perturbations were calculated at different resolutions, related by
grid refinements by a factor two in both $x$ and $\tau$.  As our
lowest resolution, we used $\Delta x=0.01$, with a Courant factor of
$\Delta\tau/\Delta x\simeq 0.05$. (The exact Courant factor is chosen so
that the number of time steps for integrating the perturbations is a
multiple of the number of time steps in the background, per period.)
This small Courant factor is necessary for stability, apparently
because some coefficients of the perturbation equations, although
smooth, have very large gradients in $\tau$ near $x=0$. We have also
verified that the effect of using an even smaller Courant factor is
negligible. Our highest resolution was finer by a factor of $16$ in
both space and time. The background coefficients $a$, $\alpha$, $X$,
$Y$ were given in Fourier coefficients at 128 points per period in
$\tau$ and the required intermediate values of $\tau$ were obtained by
local cubic interpolation. We chose local cubic interpolation as it is
much faster than Fourier interpolation, and because of limited
computer memory. The interpolation is formally second order accurate,
and all background coefficients are well resolved at this
resolution. To separate the convergence of the perturbations from that
of the background coefficients, we used a background solution obtained
with 1601 points in $x$, and downsampled it by factors of 1,
2, 4, 8 and 16.

As a test, we used all numerical methods on the trivial background of
flat empty spacetime without a scalar field. On this background, the
even matter and metric perturbations decouple. In fact, the even
matter perturbation equation is identical to the odd master equation,
and both are identical to the free wave equation.  We are therefore
testing the code on the free wave equation, with angular dependence
$Y_{lm}$, and in self-similarity coordinates, on the domain of
dependence $0\le x < 1$.

Already in this simple test, we find that the iterated Crank-Nicholson
method with any of the boundary treatments discussed is unstable. The
instability does not have a continuum limit in space. In fact, it
changes sign about every grid point in space, and grows twice as fast
in time when $\Delta x$ is halved. Nevertheless, it appears to have a
continuum limit in time.  The instability changes smoothly from one
time step to the next, and in fact, it is practically unchanged if
$\Delta \tau$ is reduced by a factor of 10 (at constant $\Delta
x$). The instability is most apparent at $x=0$, but in an implicit
scheme, all grid points in space are of course linked. 

As we have not found a boundary treatment in which iterated
Crank-Nicholson is stable, we now restrict ourselves to the two
characteristic schemes. In flat space we find that they give
essentially the same solution. Both are stable and second-order
convergent for a long time. In particular, $x=0$ and $x=1$ are
perfectly normal points, as expected from the construction of the
numerical scheme.  At high resolutions, high $l$, and late times (for
example, $\Delta x=1/1600$, $l=6$, $\tau\gtrsim 1$) we see an
oscillating instability in $\tau$ near $x=0$ that leads to a breakdown
of convergence near $x=0$, and which blows up for sufficiently large
$l$ and high resolution.  This instability appears to be a solution of
the continuum equations, but for initial data which are provided by
finite differencing error at late times. To demonstrate this, we have
evolved a narrow Gaussian pulse originally centered around $x=0.5$ in
flat spacetime. After this pulse has left the computational domain,
nothing should happen physically, and we are left with pure numerical
error. We then extracted new Cauchy data at a late time, at a high
resolution, and restarted these data with different resolutions (in
space and time), down-sampling the error data as necessary. For some
time we clearly see quadratic convergence, until new numerical error
takes over.

The instability is present already in the free wave equation in flat
space, which in our rescaled variables and self-similarity coordinates
is
\begin{eqnarray}
&& {\partial u_1\over \partial \tau} + x {\partial u_1\over\partial x}
- {\partial u_2\over\partial x} - {2l+2\over x} u_2
- (l+1) u_1 = 0, \nonumber \\
&& {\partial u_2\over \partial \tau} + x {\partial u_2\over\partial x}
- {\partial u_1\over\partial x} - (l+1) u_2 = 0.
\end{eqnarray}
Recall that for any $l$, $u_1$ is an even function of $O(1)$ of $x$
and $u_2$ is an odd function of $O(x)$.  As the instability is
centered at $x=0$ and becomes worse with increasing $l$, it must be
linked to the term $u_2/x$. We have generalized a well known trick for
the spherical wave equation which consists in absorbing this term into
the $x$-derivative. We rewrite the equations as
\begin{eqnarray}
\label{newmethod}
&& {\partial u_1\over \partial \tau} + x {\partial u_1\over\partial x}
- (2l+3){\partial (x^{2l+2} u_2)\over\partial (x^{2l+3})} 
- (l+1) u_1 = 0,  \nonumber \\
&& {\partial u_2\over \partial \tau} + x (2l+3){\partial (x^{2l+2} u_2)\over\partial (x^{2l+3})}
- {\partial u_1\over\partial x} + (l+1) u_2 = 0.
\end{eqnarray}
We then finite difference $u_2$ always in the way suggested by the
equations, using left and right second-order one-sided derivatives to
obtain the characteristic method outlined above. Note that we only
ever use the new derivative of $u_2$, never the straight derivative
$\partial u_2/\partial x$. The generalization to the even and odd
parity perturbations of scalar field collapse, applied to the
variables $u_2$ and $u_5$, is straightforward. In particular, the
coefficients of $\partial u_2/\partial x$ and $(2l+2)u_2/x$, although
now different from unity, remain equal to each other. All other
variables $u$ are differentiated directly with respect to $x$.  When
we use this finite differencing method for the flat space wave
equation, the late-time solution that is pure numerical error is now
smooth and decays exponentially instead of blowing up, as we had
hoped. On the Choptuik background, however, the instability at the
center is still not suppressed. Before the instability takes over, the
two methods clearly converge towards each other.

We have also tested convergence of the code on the (numerically
generated) Choptuik background. Here, stability requires a smaller
Courant factor. The differences between different resolutions are
peaked at $x=0$ and are smooth functions of $x$. For most values of
$\tau$ and $x$ convergence is clearly second order, with the exception
of certain value of $\tau$ near $x=0$, where the background
coefficients are particularly rapidly varying. Here convergence still
occurs (differences between resolutions decrease with resolution), but
is not of a clear order: lower than second order for low resolutions
and higher than second order for high resolutions. Somewhat
surprisingly, second-order convergence disappears and then reappears
many times.  Apparently error is not just growing with time, but
depends very strongly on the background. The explicit and iterated
characteristic methods give very similar results. At typical
resolutions the differences between the two methods at the same
resolution are much smaller than between resolutions. The only
exception from this behavior is at early times, when the iterated
method clearly shows first-order convergence that goes over smoothly
into the expected second-order convergence. 

In the flat empty background, pulses with support well inside $x=1$
quickly leave the computational domain. On the Choptuik background, we
expect to find (exponentially damped) quasiperiodic behavior at late
times.  We must therefore evolve to large values of $\tau$ (of the
order of 10 to 20 background periods $\Delta$). Not surprisingly we
find that second-order convergence breaks down after a period or so,
both because the quasiperiodic oscillations at different resolutions
drift out of phase, and because the exponential decay rates are
slightly different. At early times, we still observe second-order
convergence, as described above.



\begin{references}

\bibitem{Choptuik} M. W. Choptuik, Phys. Rev. Lett. {\bf 70}, 9
(1993).

\bibitem{Greview} C. Gundlach, Adv. Theor. Math. Phys. {\bf 2}, 1 (1998),
gr-qc/9712084. 

\bibitem{EvCol} C. R. Evans and J. S. Coleman, Phys. Rev. Lett.
{\bf 72}, 1782 (1994).

\bibitem{GMG} C. Gundlach and J. M. Mart\'\i n-Garc\'\i a, Phys.
Rev. {\bf D 54}, 7353 (1996).

\bibitem{HodPiran} S. Hod and T. Piran, Phys. Rev. D {\bf
55}, 3485  (1997). 

\bibitem{critfluid} C. Gundlach, Phys. Rev. D {\bf 57}, 7075 (1998)

\bibitem{AbrahamsEvans} A. M. Abrahams and C. R. Evans, Phys. Rev.
Lett. {\bf 70}, 2980 (1993)

\bibitem{GS1} U. H. Gerlach and U. K. Sengupta, Phys. Rev. {\bf D
19}, 2268  (1979). 

\bibitem{GS2} U. H. Gerlach and U. K. Sengupta, Phys. Rev. {\bf D
22}, 1300 (1980).

\bibitem{Gundlach} C. Gundlach, Phys. Rev. Lett. {\bf 75},
3214 (1995); Phys. Rev. D {\bf 55}, 695 (1997).

\bibitem{Zerilli} F. J. Zerilli, J. Math. Phys. {\bf 11}, 2203 (1970)

\bibitem{RW} T. Regge and J. A. Wheeler, Phys. Rev. {\bf 108}, 1063
  (1957).  

\bibitem{Leveque} R. J. LeVeque, {\it Numerical methods for
conservation laws}, Birkh\"auser, Basel 1992.


\end{references}
\end{document}